\documentclass[11pt,draftclsnofoot,journal,onecolumn]{IEEEtran}

\usepackage{graphicx}
\usepackage{subfigure}
\usepackage{amssymb}
\usepackage{algpseudocode}
\usepackage{algorithm}
\usepackage{amsmath}
\usepackage{hyperref}
\usepackage{cite}
\usepackage{cases}
\usepackage{multirow}
\usepackage{color}
\usepackage{flushend}
\usepackage{url}
\usepackage{empheq}

\IEEEoverridecommandlockouts

\newtheorem{lemma}{Lemma}

\newtheorem{theorem}{Theorem}

\newtheorem{corollary}{Corollary}

\newtheorem{observation}{Observation}
\renewcommand{\IEEEQED}{\IEEEQEDopen}

\begin{document}
\title{Primary User Traffic Classification in Dynamic Spectrum Access Networks}
\author{Chun-Hao Liu, Przemys{\l}aw Pawe{\l}czak, and Danijela Cabric%
\thanks{Chun-Hao Liu and Danijela Cabric are with the Department of Electrical Engineering, University of California, Los Angeles, 56-125B Engineering IV Building, Los Angeles, CA 90095-1594, USA (email: \{liuch37, danijela\}@ee.ucla.edu).}
\thanks{Przemys{\l}aw Pawe{\l}czak is with the Department of Electrical Engineering, Mathematics and Computer Science, Delft University of Technology, Mekelweg 4, 2628 CD Delft, The Netherlands (email: p.pawelczak@tudelft.nl).}
\thanks{This work has been supported by the National Science Foundation under CNS grant 1149981 and by the Dutch Technology Foundation STW under contract 12491.}
\thanks{Preliminary version of this work has been accepted to the proceedings of IEEE GLOBECOM, Dec. 9--13, 2013, Atlanta, GA, USA~\cite{LiuGC13}.}
\thanks{\copyright 2014 IEEE. Personal use of this material is permitted. Permission from IEEE must be 
obtained for all other uses, in any current or future media, including  reprinting/republishing this material for advertising or promotional purposes, creating new  collective works, for resale or redistribution to servers or lists, or reuse of any copyrighted  component of this work in other works.}}

\maketitle

\begin{abstract}
This paper focuses on analytical studies of the primary user (PU) traffic classification problem. Observing that the gamma distribution can represent positively skewed data and exponential distribution (popular in communication networks performance analysis literature) it is considered here as the PU traffic descriptor. We investigate two PU traffic classifiers utilizing perfectly measured PU activity (busy) and inactivity (idle) periods: (i) maximum likelihood classifier (MLC) and (ii) multi-hypothesis sequential probability ratio test classifier (MSPRTC). Then, relaxing the assumption on perfect period measurement, we consider a PU traffic observation through channel sampling. For a special case of negligible probability of PU state change in between two samplings, we propose a minimum variance PU busy/idle period length estimator. Later, relaxing the assumption of the complete knowledge of the parameters of the PU period length distribution, we propose two PU traffic classification schemes: (i) estimate-then-classify (ETC), and (ii) average likelihood function (ALF) classifiers considering time domain fluctuation of the PU traffic parameters. Numerical results show that both MLC and MSPRTC are sensitive to the periods measurement errors when the distance among distribution hypotheses is small, and to the distribution parameter estimation errors when the distance among hypotheses is large. For PU traffic parameters with a partial prior knowledge of the distribution, the ETC outperforms ALF when the distance among hypotheses is small, while the opposite holds when the distance is large.
\end{abstract}
\begin{IEEEkeywords}
Dynamic spectrum access, traffic classification, traffic sampling, traffic estimation, performance analysis.
\end{IEEEkeywords}

\IEEEpeerreviewmaketitle

\section{Introduction}
\label{sec:intro}

Dynamic/Opportunistic spectrum access (DSA/OSA) aims at increasing radio spectrum utilization~\cite{ren_wcom_2013,chai_tmc_2014}. In order to do so, the secondary (unlicensed) users (SUs) of DSA networks are allowed to transmit on licensed channels, when they are not occupied by primary (licensed) users (PUs). Understanding the PUs' channel occupancy distributions becomes important from a theoretical point of view~\cite{Cabric13}, but most importantly it allows to improve seamless DSA operation~\cite[Sec. IV-B]{kone_ton_2012},~\cite[Fig. 2]{ren_wcom_2013}. For example, if SUs have sufficient knowledge about the PUs' traffic distributions, they can minimize the channel switching latency~\cite{Shin08}, predict the PUs' behavior to minimize interference~\cite{LiuGC12} or find an optimal PU channel sensing order~\cite{liu_jsac_2013}. Therefore, the SUs should accurately estimate the PUs' traffic distribution, i.e., classify the PU traffic correctly from a set of possible distributions, e.g., exponential, gamma, log-normal, and Weibull distributions as tested in~\cite{Casadevall13}. Looking at the recent DSA/OSA applications, traffic classification can be used in Licensed Shared Access~\cite{paola_dyspan_2014} (LSA) systems, where traffic classification would help in identifying the behavior of individual LSA licensees~\cite{private_communication_2014} and adapting licensing rules accordingly.

\subsection{Related Work}
\label{sec:related_work}

Traffic classification is an important research area in many telecommunication domains, e.g. in IP networks~\cite{nguyen_survey_2008}. In parallel, analytical modeling of IP traffic has also been concerned, refer to a discussion in e.g.~\cite[Sec. III-D]{paxson_ton_1998}. In the DSA area, the topic has started to receive attention as well. Considering relevant works that aim at PU traffic classification,~\cite{Mammela10} was the first to deal with traffic pattern classification in DSA networks. Therein, the classification of the traffic pattern was done by using the autocorrelation function of the received PU signal. Work of~\cite{Mammela11} improved the classification algorithm of~\cite{Mammela10} by filtering away the errors that were caused by noise and incorrect spectrum sensing. Inspired by machine learning, the authors in~\cite{Bose12} proposed two behavior classifiers, namely a naive Bayesian classifier and an averaged one-dependence estimation classifier to classify the channel selection strategy for SUs. However, the authors of~\cite{Mammela10,Mammela11} considered the PU traffic pattern to be either stochastic or deterministic, without assigning the PU traffic to a specific distribution. Furthermore, the classifier of~\cite{Bose12} did not take the distributions of PU traffic but only the mean busy/idle time into consideration. We thus conclude, to the best of our knowledge, the performance of PU traffic classification is still relatively unexplored from the theoretical point of view.

\subsection{Our Contribution}
\label{sec:contributions}

This motivated us to perform detailed theoretical studies of PU traffic classification. Considering the classification of gamma-distributed PU busy/idle time collected through an error-free spectrum sensing process, the contribution of our work is fourfold:
\begin{enumerate}
\item We analytically derive the performance for the PU traffic classifier based on maximum likelihood using Gaussian approximation;
\item We re-evaluate a sequential algorithm based on a multi-hypothesis sequential probability ratio test~\cite{Veeravalli94}, to deal with the classification problem for multiple PU traffic classes, when parameters of PU traffic classes are known in advance;
\item Considering PU channel sampling, for a special case when probability of PU period change in-between two samples (busy-to-idle-to-busy or idle-to-busy-to-idle) is negligible, we evaluate (i) a minimum variance PU state length estimator, and (ii) propose a modified maximum likelihood classifier, quantifying its performance analytically and providing design guidelines based on traffic parameters;
\item Finally, we propose (i) a PU traffic estimate-then-classify scheme which requires no complete knowledge of the PU traffic parameters, and (ii) an average likelihood function method which requires knowledge on the statistics of the PU traffic parameters when they fluctuate in time domain.
\end{enumerate}

In addition, we list the important limitations of our work:
\begin{enumerate}
\item We assume that the set size of distributions considered for classification is finite and does not change over time;
\item The effect of spectrum sensing errors at the physical layer on the classification accuracy is not considered;
\item The calculations of classification accuracy obtained in this paper depend on the exact knowledge of a subset of traffic parameters and their stationarity.
\end{enumerate}

The rest of the paper is organized as follows. The system model is given in Section~\ref{sec:system_model}. The proposed PU traffic classifiers with perfect knowledge of PU traffic parameters are presented in Section~\ref{sec:classification_perfect}, and traffic classification using traffic period estimation schemes is presented in Section~\ref{sec:classification_blind_period}. The proposed PU traffic classifiers with imperfect knowledge of PU traffic parameters are presented in Section~\ref{sec:classification_imperfect}. Numerical results are given in Section~\ref{sec:numerical_results}. Finally, Section~\ref{sec:conclusions} concludes the paper.

\section{System Model}
\label{sec:system_model}

We consider a single channel randomly accessed by a PU. To ease the analysis we disregard (i) the effect of incidental SU operation within a PU band, i.e., the injection of SU traffic into PU traffic which obfuscates the correct classification of the latter, and (ii) the effect of spectrum sensing errors. The assumption (ii) is taken consciously, as the problem of traffic classification is strictly coupled with the spectrum sensing problem and requires a separate analytical study due to its complexity. For example, in~\cite[Sec. 4.2]{Wellens_monet_2010} it has been concluded that ``different energy detection thresholds (\dots) result in significantly different [PU traffic] distributions.'' Recent work of~\cite{Benitez13} provides a more formal discussion on the effect of sensing errors on PU traffic analysis. Nevertheless, assumptions (i) and (ii) allow us to use the results obtained in this paper also for the non-DSA scenarios and provide a classification benchmark for interference-prone and sensing error-prone cases.

Further, we assume we can obtain traffic busy/idle periods (denoted as ON/OFF, respectively) perfectly through time-domain fine-grained spectrum sensing, as in e.g.~\cite[Sec. II]{Zhao12}. This assumption, in practical terms, results in a sampling time much smaller than the shortest duration of PU traffic periods. The ON/OFF periods are denoted as a random variable $X$ with its $n$ independent and identically distributed realizations $\mathbf{x}=(x_1,x_2,\cdots,x_n)^T,~x_i\in(0,\infty)$. Those are assumed to belong to one of $\mathcal{M}=\{1,\cdots,M\}$ possible gamma distributions. The gamma distribution is chosen for its flexibility to represent: (i) exponential distribution, due to its analytical popularity~\cite[Sec. V-B]{jung_ton_2012} and existence in real networks, e.g. as measured in~\cite[Sec. IV-A]{wilkomm_dyspan_2008} for call arrival times in CDMA-based system; and (ii) positively skewed data, which is also confirmed through the traffic measurement, e.g. in~\cite[Fig. 10]{sharp_jsac_2004} for call holding time in public safety systems.

Our objective is to minimize the required number of measurement periods in $\mathbf{x}$ in order to classify $X$ to the correct distribution. We can formulate such a classification problem as a multi-hypothesis problem, i.e.,
\begin{align}
X\sim f_X(x)=\begin{cases}
f_1(x|\mathbf{\Theta}_1), & \quad \mathcal{H}_1,\\
f_2(x|\mathbf{\Theta}_2), & \quad \mathcal{H}_2,\\
\qquad \vdots & ~ \\
f_M(x|\mathbf{\Theta}_M), & \quad \mathcal{H}_M,
\end{cases}
\end{align}
where $f_X(x)$ is the hypothesized probability density function (PDF) of $X$, $f_j(x|\mathbf{\Theta}_j)=\frac{\beta_j^{\alpha_j}}{\Gamma(\alpha_j)}x^{\alpha_j-1}e^{-\beta_j x}$ is the gamma PDF of $X$ under hypothesis $\mathcal{H}_j$ given the shape parameter $\alpha_j$ and the rate parameter $\beta_j$, where $\mathbf{\Theta}_j=(\alpha_j,\beta_j)^T$ and $\Gamma(x)=\int_0^{\infty}t^{x-1}e^{-t}dt$ is the gamma function, where again $x\in(0,\infty)$. We assume that each hypothesis $\mathcal{H}_j$ has a prior probability $\pi_j$, and we define $\mathbf{\Omega}=(\pi_1,\pi_2,\cdots,\pi_M)^T$, $\sum\limits_{i=1}^{M}\pi_i=1$. Without loss of generality, in this paper we assume that the elements in $\mathbf{x}$ denote either PU channel occupancy periods (ON times) or idle periods only (OFF times).

\section{Traffic Classification with Perfect Knowledge of PU Traffic Parameters}
\label{sec:classification_perfect}

We start with assuming a perfect knowledge of all PU traffic parameters $\mathbf{\Theta}_j=(\alpha_j,\beta_j)^T$, $\forall j\in\mathcal{M}$. Firstly, we introduce a maximum likelihood classifier (MLC) that requires a constant number of PU traffic periods, which is an optimal classifier in terms of probability of correct classification when the PDFs are known~\cite[Sec. I]{Dobre07} and derive its classification performance for the considered model in Section~\ref{sec:system_model}. Such an analysis, to the best of our knowledge, has not been performed before. Secondly, as a comparison to MLC, we re-introduce the multi-hypothesis sequential probability ratio test classifier (MSPRTC) using~\cite{Veeravalli94} which adopts a sequential sample test instead of using a fixed number of PU traffic periods for classification.

\subsection{Maximum Likelihood Classifier}
\label{sec:MLC}

For the considered gamma distribution $f_j(x|\mathbf{\Theta}_j)$ the likelihood function given $\mathbf{x}$ for $\mathcal{H}_j$ can be written as
\begin{align}
L_{\mathcal{H}_j}(\mathbf{x})&=\pi_j\prod\limits_{i=1}^{n}f_j(x_i|\mathbf{\Theta}_j)\nonumber\\
&=\pi_j\prod\limits_{i=1}^{n}\left(\frac{\beta_j^{\alpha_j}}{\Gamma(\alpha_j)}x_i^{\alpha_j-1}e^{-\beta_j x_i}\right), \forall j\in\mathcal{M}.
\label{eq:LF}
\end{align}
Then, the MLC final decision, $\nu$, is
\begin{equation}
\nu=\mathcal{H}_{m\triangleq \arg\max\limits_j L_{\mathcal{H}_j}(\mathbf{x})}.
\label{eq:MLC}
\end{equation}

To analyze the MLC classification performance for the system model considered in Section~\ref{sec:system_model}, we start with calculating the log-likelihood function $g_{\mathcal{H}_j}(\mathbf{x})\triangleq\log L_{\mathcal{H}_j}(\mathbf{x})$ which can be represented as
\begin{align}
g_{\mathcal{H}_j}(\mathbf{x})=\log\frac{\pi_j\beta_j^{n\alpha_j}}{\Gamma(\alpha_j)^n}+\sum\limits_{i=1}^{n}[(\alpha_j-1)\log x_i-\beta_j x_i].
\label{eq:gj}
\end{align}
Then we can calculate the probability of correct classification under $\mathcal{H}_j$ using~(\ref{eq:gj}) as
\begin{align}
\Pr\{\nu=\mathcal{H}_j|&\mathcal{H}_j\}
=\Pr\{g_{\mathcal{H}_j}(\mathbf{x})>g_{\mathcal{H}_k}(\mathbf{x})\}~\forall k\in\{\mathcal{M}-\{j\}\}\notag\\
&=\prod_{k=1,~k\neq j}^M\Pr\{g_{\mathcal{H}_j}(\mathbf{x})-g_{\mathcal{H}_k}(\mathbf{x})>0\}.
\label{eq:cond1}
\end{align}
Embedding~(\ref{eq:gj}) into~(\ref{eq:cond1}) we can simplify (\ref{eq:cond1}) as
\begin{align}
&\Pr\{\nu=\mathcal{H}_j|\mathcal{H}_j\}\nonumber\\&=\prod_{k=1,~k\neq j}^M\Pr\left\{\sum\limits_{i=1}^{n}y_i^{(j,k)}>-\log\frac{\pi_j\beta_j^{n\alpha_j}\Gamma(\alpha_k)^n}{\pi_k\beta_k^{n\alpha_k}\Gamma(\alpha_j)^n}\right\},
\label{eq:cond2}
\end{align}
where $y_i^{(j,k)}=\alpha_{j,k}\log x_i-\beta_{j,k}x_i$ and $\alpha_{j,k}=\alpha_j-\alpha_k$, $\beta_{j,k}=\beta_j-\beta_k$. We also define the mean and variance for the variable $y_i^{(j,k)}$ as $\mu_{j,k}$ and $\sigma_{j,k}^2$, respectively, which are derived in Appendix~\ref{sec:pdf}.

We can now define $\bar y^{(j,k)}\triangleq\sum\limits_{i=1}^{n}y_i^{(j,k)}$ and calculate its PDF as $f\left(\bar y^{(j,k)}\right)=f^{(n)}\left(y_i^{(j,k)}\right)$, where $f^{(n)}(\cdot)$ denotes the $n$-fold PDF convolution. Then, by calculating the cumulative distribution function (CDF) of $\bar y^{(j,k)}$ we can obtain an exact analytical expression for~(\ref{eq:cond2}). However, due to mathematical intractability of such operations we use a simple approximation instead, which has a closed-form expression, to derive the probability of correct classification. Therefore, let us transform~(\ref{eq:cond2}) as
\begin{align}
\Pr\{\nu=\mathcal{H}_j|\mathcal{H}_j\}=\prod_{k=1,~k\neq j}^M\Pr\{z_{j,k}>\tau_{j,k}\},
\label{eq:cond3}
\end{align}
where $z_{j,k}=\frac{1}{\sqrt{n\sigma_{j,k}^2}}\sum\limits_{i=1}^n\left(y_i^{(j,k)}-\mu_{j,k}\right)$ and $\tau_{j,k}=-\frac{1}{\sqrt{n\sigma_{j,k}^2}}\left(\log\frac{\pi_j\beta_j^{n\alpha_j}\Gamma(\alpha_k)^n}{\pi_k\beta_k^{n\alpha_k}\Gamma(\alpha_j)^n}+n\mu_{j,k}\right)$. According to the Central Limit Theorem, as $n$ is large enough, $z_{j,k}$ will approach a standard normal distribution, $\mathcal{N}(0,1)$. Hence we can approximate (\ref{eq:cond3}) as
\begin{align}
\Pr\{\nu=\mathcal{H}_j|\mathcal{H}_j\}\approx\prod_{k=1,~k\neq j}^M Q(\tau_{j,k}),
\label{eq:cond4}
\end{align}
where $Q(\cdot)$ is the tail probability function of the standard normal distribution. Finally, the average probability of correct classification $P_c$ for all hypotheses is derived using (\ref{eq:cond4}) as
\begin{empheq}[box=\fbox]{align}
P_c=\sum\limits_{j=1}^M\pi_j\Pr\{\nu=\mathcal{H}_j|\mathcal{H}_j\}.
\label{eq:Pc}
\end{empheq}

\subsection{Multi-Hypothesis Sequential Probability Ratio Test Classifier}
\label{sec:MSPRT}

To compare the performance with MLC, we introduce a new classification method based on MSPRTC of~\cite{Veeravalli94}. Unlike MLC which uses a constant number of PU traffic ON (or OFF) periods, MSPRTC sequentially classifies multiple hypotheses requiring only as many PU traffic periods as needed for correct classification. We adopt MSPRTC since the authors in~\cite[Sec. III]{Veeravalli94} show that it provides a good approximation to the optimal solution on the condition of a perfect a priori knowledge for all distributions, i.e. their parameters, in the sequential multi-hypothesis classification problem.

MSPRTC decision is then $\nu=\mathcal{H}_{m\triangleq \arg\max\limits_j p_{N_A}^j}$, where the posteriori probability $p_n^j$ is given as~\cite[Sec. II]{Veeravalli94} 
\begin{equation}
p_n^j\triangleq{\pi_j\prod\limits_{i=1}^{n}f_j(x_i|\mathbf{\Theta}_j)}{\left[\sum\limits_{l=1}^{M}\pi_l\left(\prod\limits_{i=1}^{n}f_l(x_i|\mathbf{\Theta}_l)\right)\right]^{-1}}.
\label{eq:pos}
\end{equation}
We define $N_A$ as the first $n\geq 1$ such that $p_n^j>\frac{1}{1+A_j}$ for at least one $j\in\mathcal{M}$, where $A_j>0$ is the design threshold.

Recalling~\cite[Sec. VII]{Veeravalli94} $A_k=\frac{c}{\delta_k\gamma_k}$, where $c=\frac{\alpha}{\sum_{k=0}^{M-1}\frac{\pi_k}{\delta_k}}$, $\alpha$ is the total probability of incorrect decision, $\gamma_k$ is the constant defined in~\cite[Sec. VI]{Veeravalli94} and $\delta_k$ is the measure of probabilistic distance. In~\cite[Sec. VII]{Veeravalli94} $\delta_k=\min\limits_{k:k\neq j}D(f_j(x|\mathbf{\Theta}_j),f_k(x|\mathbf{\Theta}_k))$, $D$ is the Kullback-Leibler (KL) divergence which for two gamma distributions is defined in~\cite[Eq. (6)]{Penny01} and after simplifications
\begin{align}
&D(f_j(x|\mathbf{\Theta}_j),f_k(x|\mathbf{\Theta}_k))\triangleq\int_{-\infty}^{\infty}\!f_j(x|\mathbf{\Theta}_j)\log\frac{f_j(x|\mathbf{\Theta}_j)}{f_k(x|\mathbf{\Theta}_k)}dx\nonumber\\&=(\alpha_j-\alpha_k)\psi(\alpha_j)-\log\Gamma(\alpha_j)\notag\\&+\log\Gamma(\alpha_k)+\alpha_k(\log\beta_k-\log\beta_j)+\alpha_j\left(\frac{\beta_j-\beta_k}{\beta_k}\right),
\label{eq:KLD}
\end{align}
where $\psi(x)=\frac{\Gamma'(x)}{\Gamma(x)}$ is the digamma function\footnote{For the derivation see \url{http://stats.stackexchange.com/questions/11646/kullbackleibler-divergence-between-two-gamma-distributions}, retrieved December 22, 2013.}

\begin{observation}
The authors of~\cite{Veeravalli94} suggest to use KL for $\delta_k$ as a descriptor of probabilistic distance for two distributions. 
For the squared Hellinger (SH) distance, defined as~\cite[Ch. 14.5, pp. 211]{vandervaart_cup_2002}
\begin{align}
H^2(f_j(x|\mathbf{\Theta}_j),&f_k(x|\mathbf{\Theta}_k))\nonumber\\&\triangleq1-\int_{-\infty}^{\infty}\sqrt{f_j(x|\mathbf{\Theta}_j)f_k(x|\mathbf{\Theta}_k)}dx,
\label{eq:SHD}
\end{align}
(note that the 0.5 constant is omitted for convenience as remarked in~\cite[Ch. 3.3, pp. 61]{pollard_cup_2002}), it can be shown to be the lower bound of KL divergence~\cite[Proposition 1]{castro_lecture_2013}, i.e.,
\begin{align}
D(f_j(x|\mathbf{\Theta}_j), f_k(x|\mathbf{\Theta}_k))\geq H^2(f_j(x|\mathbf{\Theta}_j), f_k(x|\mathbf{\Theta}_k)).
\end{align}
We thus propose to replace $\delta_i$ used in calculating the threshold for MSPRTC, $A_j$, with $\eta_j$ where 
\begin{align}
\eta_j=\min\limits_{k:k\neq j}H^2(f_j(x|\mathbf{\Theta}_j),f_k(x|\mathbf{\Theta}_k)),
\end{align} 
\end{observation}
and the SH distance between two gamma distributions (considered in the system model in Section~\ref{sec:system_model}) is derived in Appendix~\ref{sec:SH_2_Gamma}.

\begin{observation}
The procedure to calculate $\gamma_k$ explained in~\cite[Sec. VII]{Veeravalli94} is convolved\footnote{Even though we used it in~\cite{LiuGC13} by actually not calculating it, but sweeping through a large set of values of constant $\gamma$ (Bayes classification risk minimizer) to obtain a desired classification.}. Therefore, in numerical evaluation in Section~\ref{sec:numerical_results} we will replace $A_k$ with a single value $\gamma$ for all the hypotheses. To find $\gamma$, before performing classification we sweep through $\gamma\in[0,\infty)$ to determine the desired classification probability. For example, we can set $\gamma=0$ and obtain the first classification performance. If it does not satisfy the classification system requirement, we increase $\gamma$ by a pre-defined step size $\Delta \gamma>0$ until we reach our desired classification performance.
\end{observation}

\section{Joint PU Traffic Period Estimation and Traffic Classification}
\label{sec:classification_blind_period}

So far, we have assumed the continuous observation of the PU channel state. In this section we consider a more general traffic classification problem, where the elements of $\mathbf{x}$ also need to be estimated. Therefore we relax the assumption on the continuous observation of PU state and assume a PU channel observation at instants every $T_s$ seconds to find the elements in $\mathbf{x}$.

First, we introduce the model for the PU period length estimation in Section~\ref{sec:noise_modeling}. Then, in Section~\ref{sec:estimation_sampling}, we propose a minimum variance period length estimator to minimize estimation errors. Subsequently, we propose a modified MLC considering estimation error and analytically derive the approximation of its classification performance in Section~\ref{sec:MLC_Error}. We then propose a modified MSPRTC considering estimation error in Section~\ref{sec:MSPRT_Error}. Finally, in Section~\ref{sec:Guideline} we propose a design guideline for MLC with energy or time constraints on the spectrum sensing budget.

\subsection{Period Estimation Noise Modeling under PU Traffic Sampling}
\label{sec:noise_modeling}

We follow the system model shown in~\cite[Section II, Fig. 1(a)]{Benitez13}, where a PU traffic period, i.e., ON/OFF duration $T_{\text{on}}$/$T_{\text{off}}$, is estimated through sampling performed at regular intervals of $T_s$ seconds. Without loss of generality, we will focus on estimating $T_{\text{on}}$ only, while $T_{\text{off}}$ can be estimated using the same technique. In addition, to ease the analysis, we assume that the probability of PU state change between two samplings is negligible.

Denote $s=1$ represents the channel being busy, while $s=0$ represents the channel being idle. Assuming as previously that we ignore spectrum sensing errors we would like to estimate the length of $T_{\text{on}}$ based on the set of samples obtained at $T_s$ intervals. For the actual $T_{\text{on}}$ we denote four time instants, i.e., $t_0$, $t_1$, $t_2$, and $t_3$: (i) $t_0$ is the starting point with $s=0$, (ii) $t_1$ and (iii) $t_2$ are the transition points from $s=0$ to $s=1$ and $s=1$ to $s=0$, respectively, and (iv) $t_3$ is the end point with $s=0$. After sampling the traffic, we define the nearest sampling point to $t_1$ as $\zeta_1$ in region $(t_0,t_1)$ and $\zeta_2$ in region $(t_1,t_2)$. Similarly, we define the nearest sampling point to $t_2$ as $\zeta_3$ in region $(t_1,t_2)$ and $\zeta_4$ in region $(t_2,t_3)$. In other words, $\zeta_i$ are the actual discrete channel measurement points. Then we can think of this PU channel sampling as a quantization process, i.e., there are four sources of quantization noise which are $\phi_1=t_1-\zeta_1$, $\phi_2=\zeta_2-t_1$, $\phi_3=t_2-\zeta_3$, and $\phi_4=\zeta_4-t_2$. We can now model quantization error as a uniformly distributed random variable, which implies that $\phi_i\thicksim\mathcal{U}(0,T_s),~\forall i\in\{1,2,3,4\}$, where $\mathcal{U}(a,b)$ denotes the uniform distribution and $a$, $b$ are the minimum and maximum value for the random variable $\phi_i$, respectively.

\subsection{$T_{\text{on}}$ Length Estimator}
\label{sec:estimation_sampling}

We first propose a minimum variance PU period length estimator that reduces the sampling noise effect. Then we derive the average number of PU traffic samples needed for $T_{\text{on}}$ length estimation using the proposed estimator.

\subsubsection{Minimum Variance Estimator}
\label{sec:minimum_variance_estimator}

First we consider $T_1$, i.e., the interval between two nearest $s=0$ points, where $T_1=\zeta_4-\zeta_1=T_{\text{on}}+\phi_1+\phi_4$. Then, we consider $T_2$, i.e., the interval between two nearest $s=1$ points, where $T_2=\zeta_3-\zeta_2=T_{\text{on}}-\phi_2-\phi_3$. We propose a weighted average of $T_1$ and $T_2$, i.e., $T_a=wT_1+(1-w)T_2$ as our $T_{\text{on}}$ estimator, where $w\in[0,1]$ is the weight that needs to be designed. We know that the mean for $T_a$ is 
\begin{align}
\mathbb{E}\{T_a\}
&=w\mathbb{E}\{T_1\}+(1-w)\mathbb{E}\{T_2\}\nonumber\\&=w\mathbb{E}\{T_{\text{on}}+\phi_1+\phi_4\}+(1-w)\mathbb{E}\{T_{\text{on}}-\phi_2-\phi_3\}\notag\\
&=(2w-1)T_s+\mathbb{E}\{T_{\text{on}}\},
\end{align}
since $\mathbb{E}\{\phi_i\}=\frac{T_s}{2},~\forall i\in\{1,2,3,4\}$. We can observe that with $w=\frac{1}{2}$, the mean of $T_a$ will be the same as the mean of $T_{\text{on}}$, resulting in $T_a$ an unbiased estimator. Then we would like to minimize the variance of $T_a$ to derive the optimal $w$. Such variance is expressed as
\begin{align}
\text{Var}\{T_a\}=\frac{T_s^2}{6}(w^2+(1-w)^2)+\text{Var}\{T_{\text{on}}\},
\label{eq:variance}
\end{align}
since $\text{Var}\{\phi_i\}=\frac{T_s^2}{12},~\forall i\in\{1,2,3,4\}$. Taking the derivative of~(\ref{eq:variance}) with respect to $w$ and setting it to zero, we can obtain the optimal weight as $w^{*}=\frac{1}{2}$. Therefore, the minimum variance estimator (MVE) is expressed as
\begin{align}
T_a=\frac{1}{2}(T_1+T_2)=T_{\text{on}}+\phi_1-\phi_3=T_{\text{on}}-\phi_2+\phi_4.
\label{eq:Ton2}
\end{align}
\subsubsection{Average Number of PU Traffic Samples for Period Estimation using Minimum Variance Estimator}
\label{sec:average_number}

The following theorem summarizes the analytical results for the average number of PU traffic samples, $N$, when we adopt the proposed MVE to estimate one PU state length $T_{\text{on}}$. 
\begin{theorem}
The expected average number of traffic samples for estimating one PU period is
\begin{empheq}[box=\fbox]{align}
\mathbb{E}\{N\}=\sum\limits_{j=1}^{M}\pi_j\mathbb{E}\{N|\mathcal{H}_j\},
\label{eq:N}
\end{empheq}
where 
\begin{align}
\mathbb{E}\{N|\mathcal{H}_j\}=\sum\limits_{k=1}^{\infty}\frac{\Gamma(\alpha_j,k\beta_jT_s)}{\Gamma(\alpha_j)}+1.
\label{eq:N_Hj}
\end{align}
\end{theorem}
\begin{IEEEproof}
See Appendix~\ref{sec:samples_sampling}.
\end{IEEEproof}
\begin{corollary}
If hypothesis $\mathcal{H}_j$ is an exponential distribution with parameter $\lambda$, then
\begin{align}
\mathbb{E}\{N|\mathcal{H}_j\}=\frac{1}{1-e^{\lambda T_s}}.
\end{align} 
\end{corollary}
\begin{IEEEproof}
We can simplify~(\ref{eq:N_Hj}) by assigning $\alpha_j=1$ and $\beta_j=\lambda$, which results in
\begin{align}
\mathbb{E}\{N|\mathcal{H}_j\}&=\sum\limits_{k=1}^{\infty}\frac{\Gamma(1,k\lambda T_s)}{\Gamma(1)}+1=\sum\limits_{k=1}^{\infty}\int_{k\lambda T_s}^{\infty}e^{-t}dt+1\nonumber\\
&=\sum\limits_{k=1}^{\infty}e^{-k\lambda T_s}+1=\frac{1}{1-e^{-\lambda T_s}}.
\label{eq:coro1}
\end{align}
\end{IEEEproof}

\begin{corollary}
If hypothesis $\mathcal{H}_j$ is an Erlang distribution with parameter $\alpha_j=2$ and $\beta_j=\lambda$ then
\begin{align}
\mathbb{E}\{N|\mathcal{H}_j\}=\frac{1-e^{-\lambda T_s}+\lambda T_se^{-\lambda T_s}}{(1-e^{-\lambda T_s})^2}.
\end{align} 
\end{corollary}
\begin{IEEEproof}
If $\alpha_j$ is an integer then $\Gamma(\alpha_j)=(\alpha_j-1)!$ and $\Gamma(\alpha_j,k\lambda T_s)=(\alpha_j-1)!e^{-k\lambda Ts}\sum\limits_{l=0}^{\alpha_j-1}\frac{(k\lambda T_s)^l}{l!}$. Plugging the above two results with $\alpha_j=2$ into~(\ref{eq:N_Hj}) we have
\begin{align}
\mathbb{E}\{N|\mathcal{H}_j\}&=\sum\limits_{k=1}^{\infty}e^{-k\lambda T_s}\sum\limits_{l=0}^{1}\frac{(k\lambda T_s)^l}{l!}+1\nonumber\\&=\left(\sum\limits_{k=1}^{\infty}e^{-k\lambda T_s}+1\right)+\sum\limits_{k=1}^{\infty}k\lambda T_se^{-k\lambda T_s}.
\label{eq:coro2}
\end{align}
The left hand side in~(\ref{eq:coro2}) can be simply obtained from~(\ref{eq:coro1}), and the right hand part in~(\ref{eq:coro2}) can be calculated as $\sum\limits_{k=1}^{\infty}k\lambda T_se^{-k\lambda T_s}=\frac{\lambda T_se^{-\lambda T_s}}{(1-e^{-\lambda T_s})^2}$. Combining the left hand and right hand parts completes the proof.
\end{IEEEproof}
\subsection{MLC under PU Period Estimation Error}
\label{sec:MLC_Error}

To derive the MLC considering PU period estimation error, we first need to derive the modified PDF for our proposed estimator. From~(\ref{eq:Ton2}) we can observe that the estimated PU period length is represented by the real PU traffic periods plus two uniformly distributed variables (representing sampling noise), one for the beginning and one for the end of the PU traffic period. The PDF for the combined sampling noise, $\phi=\phi_1-\phi_3$ or $\phi=-\phi_2+\phi_4$, can be calculated by taking the convolution of two uniform distributions, which can be expressed as a triangular function
\begin{align}
f_{\Phi}(\phi)=\Lambda(-T_s,T_s)=\frac{I(-\phi)}{T_s^2}\phi+\frac{1}{T_s},
\label{eq:noisePDF}
\end{align}
where $I(\phi)=1$ if $\phi\geq0$, else $I(\phi)=-1$. By convolving the PDF for $T_{\text{on}}$ and $\phi$, we can obtain the PDF for $T_a$, which can be derived using the following theorem.
\begin{theorem}
Given a random variable $\tilde x=x+\phi$, where $x$ is gamma distributed with parameters $\mathbf{\Theta}=(\alpha,\beta)$ and $\phi$ is triangular distributed with parameter $T_s$, the PDF of $\tilde x$ can be expressed as
\begin{align}
&f(\tilde x|\mathbf{\Theta},T_s)=\begin{cases}
\frac{\Gamma(\alpha+1,(\tilde x+T_s)\beta)-2\Gamma(\alpha+1,\tilde x\beta)+\Gamma(\alpha+1,(\tilde x-T_s)\beta)}{\Gamma(\alpha)\beta T_s^2}\\
\quad-\frac{(\tilde x+T_s)\Gamma(\alpha,(\tilde x+T_s)\beta)}{\Gamma(\alpha) T_s^2}\\
+\frac{2 \tilde x\Gamma(\alpha,\tilde x\beta)-(\tilde x-T_s)\Gamma(\alpha,(\tilde x-T_s)\beta)}{\Gamma(\alpha) T_s^2}, & \!\!\!\!\!\!\!\!\!\!\!\!\!\!\!\!\!\!\!\!\text{if}~\tilde x\geq T_s,\\
\frac{\Gamma(\alpha+1,(\tilde x+T_s)\beta)-2\Gamma(\alpha+1,\tilde x\beta)+\Gamma(\alpha+1)}{\Gamma(\alpha)\beta T_s^2}\\
\quad-\frac{(\tilde x+T_s)\Gamma(\alpha,(\tilde x+T_s)\beta)}{\Gamma(\alpha) T_s^2}\\
\quad+\frac{2 \tilde x\Gamma(\alpha,\tilde x\beta)-(\tilde x-T_s)\Gamma(\alpha)}{\Gamma(\alpha) T_s^2}, & \!\!\!\!\!\!\!\!\!\!\!\!\!\!\!\!\!\!\!\!\text{if}~0\leq \tilde x<T_s,\\
\frac{\Gamma(\alpha+1,(\tilde x+T_s)\beta)-\Gamma(\alpha+1)}{\Gamma(\alpha)\beta T_s^2}\\
\quad-\frac{(\tilde x+T_s)[\Gamma(\alpha,(\tilde x+T_s)\beta)-\Gamma(\alpha)]}{\Gamma(\alpha) T_s^2}, & \!\!\!\!\!\!\!\!\!\!\!\!\!\!\!\!\!\!\!\!\text{if}-T_s\leq \tilde x<0,\\
0,& \!\!\!\!\!\!\!\!\!\!\!\!\!\!\!\!\!\!\!\!\text{otherwise}.
\end{cases}
\label{eq:mPDF}
\end{align}
\label{theorem:y}
\end{theorem}
\begin{IEEEproof}
See Appendix~\ref{sec:likelihood_noise}.
\end{IEEEproof}
Denote the realization for $T_a$ as $\tilde x_i$. We can obtain its PDF, $f_j(\tilde x_i|\mathbf{\Theta},T_s)$, under hypothesis $\mathcal{H}_j$ from Theorem~\ref{theorem:y}. We follow the same step in Section~\ref{sec:MLC} to derive the MLC, where the likelihood function can be written as $L_{\mathcal{H}_j}(\mathbf{\tilde x})=\pi_j\prod\limits_{i=1}^{n}f_j(\tilde x_i|\mathbf{\Theta},T_s), \forall j\in\mathcal{M}$, similarly to (\ref{eq:LF}).

To quantify the probability of correct classification with estimation error, $\tilde P_c$, in a closed-form, we apply approximation in the same manner as in Section~\ref{sec:MLC}. First, let us assume that the sampling period $T_s$ is not large, which means that PDF of $\tilde x_i$ and $x_i$ would not significantly deviate from each other. We first replace $y_i^{(j,k)}$ with the $\tilde y_i^{(j,k)}=\alpha_{j,k}\log (x_i+\phi_i)-\beta_{j,k}(x_i+\phi_i)$ where $\phi_i$ is a realization for the quantization noise. To be able to apply (\ref{eq:cond2}) considering sampling noise we need to first find an expectation and variance of $\tilde y_i^{(j,k)}$, i.e., $\tilde\mu_{j,k}$ and $\tilde\sigma_{j,k}^2$, respectively.  For $\tilde\mu_{j,k}$, since $x_i+\phi_i$ might be a negative value, the mean for $\tilde y_i^{(j,k)}$ might be a complex number, which can not be used in the Q function. Therefore we use $\tilde\mu_{j,k}=\mu_{j,k}$. On the other hand, the derivation for $\tilde\sigma_{j,k}^2$ is given in Appendix~\ref{sec:new_statistics}, which is always a real number. We can now obtain the average probability of correct classification $\tilde P_c$ under estimation noise using~(\ref{eq:cond4}) and~(\ref{eq:Pc}) by replacing $\sigma_{j,k}^2$ with $\tilde\sigma_{j,k}^2$. It is thus imperative to emphasize that the proposed calculation method (due to above assumptions) is quite inaccurate considering all parameter combinations and needs to be taken with caution. Therefore calculation of classification performance is still considered to be an open problem. The reader is encouraged to experiment with our analytical procedure of classification based on the accompanying MATLAB code, see Section~\ref{sec:result_reproducibility}.

\subsection{MSPRTC under Period Estimation Error}
\label{sec:MSPRT_Error}

The proposed MSPRTC under period estimation error follows the same procedure explained in Section~\ref{sec:MSPRT}. The only adaptation is to replace the PDF $f_j(x|\mathbf{\Theta}_j)$ in~(\ref{eq:pos}) with the modified PDF $f_j(\tilde x|\mathbf{\Theta}_j,T_s)$ derived in~(\ref{eq:mPDF}).

\subsection{A Design Guideline for Traffic Classification using MLC}
\label{sec:Guideline}

There are two parameters, i.e., total observation time, $T$, and total number of samples, $N$, to be used in classification that need to be optimized. Naturally, we would like to use the smallest $T$ or $N$ to achieve the desired performance for MLC. To derive the performance of correct classification using period estimation $\tilde P_c$, we need to obtain the number of periods and the sampling period $T_s$. Obviously $T_s=\frac{T}{N-1}$. The average number of periods can be derived as
\begin{align}
\mathbb{E}\{K\}=\sum\limits_{j=1}^M \pi_j\mathbb{E}\{K|\mathcal{H}_j\},
\label{eq:K}
\end{align}
where
\begin{align}
\mathbb{E}\{K|\mathcal{H}_j\}=\frac{T}{\mathbb{E}\{T_{\text{on}}|\mathcal{H}_j\}}(1-R(T_s|\mathcal{H}_j)).
\label{eq:Ki}
\end{align}
Here $\mathbb{E}\{K|\mathcal{H}_j\}$ is the average number of periods we can obtain under hypothesis $\mathcal{H}_j$, which is equal to the total average number of periods $\frac{T}{\mathbb{E}\{T_{\text{on}}|\mathcal{H}_j\}}$ times the successful period detection rate $1-R(T_s|\mathcal{H}_j)$, where $R$ is the mis-detection rate for detecting one period defined as $R(T_s|\mathcal{H}_j)=\Pr\{T_{\text{on}}<T_s|\mathcal{H}_j\}=G(T_s|\mathbf{\Theta}_j)$, where $G(\cdot|\mathbf{\Theta}_j)$ is the CDF function for a gamma distribution under hypothesis $\mathcal{H}_j$. Note that $T_s$ and $\mathbb{E}\{K\}$ are functions of $T$ and $N$, therefore we know $\tilde P_c$ is a function of traffic parameters $\mathbf{\Theta}_j,~\forall j\in\mathcal{M}$, $\mathbf{\Omega}$, observation time $T$, and number of traffic samples $N$. Once the classification performance constraint $\epsilon$ is given, we can solve the optimization problem 
\begin{equation}
\min~T(\text{or}~N) \text{ subject to } \tilde P_c\geq\epsilon
\end{equation}
analytically.

\section{Traffic Classification with Imperfect Knowledge of PU Traffic Parameters}
\label{sec:classification_imperfect}

We further relax the system model assumptions from Section~\ref{sec:system_model} and consider the lack of complete information on $\mathbf{\Theta}_j$. Specifically, for the perfectly measured $\mathbf{x}$ we assume that the shape parameters $\alpha_j$ are known, but the rate parameters $\beta_j$, $\forall j\in\mathcal{M}$ are not. 

First, we consider to treat $\beta_j$ as unknown deterministic value. In this case we propose the \emph{estimate-then-classify} (ETC) scheme to complete the traffic classification, where we estimate all PU traffic parameters before applying them to the MLC (Section~\ref{sec:ETC_MLC}) and MSPRTC (Section~\ref{sec:ETC_MSPRT}). Additionally, for the ETC we derive the classification performance of MLC analytically. Then, if the PU traffic parameters $\beta_j$ follow a certain distribution, we propose in Section~\ref{sec:classification_prior} the \emph{average likelihood function} (ALF) for the classifiers.

\subsection{Estimate-Then-Classify: Using MLC}
\label{sec:ETC_MLC}

The ML estimator of $\beta_j$ for the distribution $f_j(x|\mathbf{\Theta}_j)$ can be derived by solving $\frac{\partial \prod\limits_{i=1}^{n}f_j(x_i|\mathbf{\Theta}_j)}{\partial\beta_j}=0$ which gives
\begin{equation}
\hat\beta_j={\alpha_j n}{\left(\sum\limits_{i=1}^{n}x_i\right)^{-1}}.
\label{eq:MLE}
\end{equation}
Considering MLC, the ETC scheme is based on replacing $\beta_j$ with its estimate $\hat\beta_j$ in the PDF of $x$ as $f_j(x|\mathbf{\hat\Theta}_j)$ where $\mathbf{\hat\Theta}_j=(\alpha_j,\hat\beta_j)$ and subsequently to the likelihood function defined in (\ref{eq:LF}).

To analyze the classification performance for the proposed ETC-based MLC, we can simply use~(\ref{eq:Pc}) except $\beta_j$ is replaced by the corresponding mean of the estimator $\mathbb{E}\{\hat \beta_j\}$. To be more specific, under hypothesis $\mathcal{H}_j$, the mean is expressed as $\mathbb{E}\{\hat\beta_k^{-1}|\mathcal{H}_j\}=\frac{1}{n\alpha_k}\sum\limits_{i=1}^{n}\mathbb{E}\{x_i|\mathcal{H}_j\}=\frac{\alpha_j}{\alpha_k}\beta_j^{-1}$, and the variance is expressed as $\text{Var}\{\hat\beta_k^{-1}|\mathcal{H}_j\}=\frac{\alpha_j\beta_j^{-2}}{n\alpha_k^2}$, $\forall k\in\mathcal{M}$. Therefore as $n$ approaches infinity, the variance for $\hat\beta_k^{-1}$ approaches zero, which means that $\hat\beta_k^{-1}$ converges to $\frac{\alpha_j}{\alpha_k}\beta_j^{-1}$ asymptotically. To conclude, we replace $\beta_k$ with $\mathbb{E}\{\hat\beta_k\}$ and embed it into~(\ref{eq:Pc}), and the probability of correct classification using ETC-based MLC can be represented as
\begin{empheq}[box=\fbox]{align}
&\hat{P_c}
=\sum\limits_{j=1}^M\pi_j\prod_{k=1, k\neq j}^M Q\left(-\frac{1}{\sqrt{n\hat\sigma_{j,k}^2}}\right.\nonumber\\&\left.\quad\times\left(\log\frac{\pi_j\alpha_j^{n\alpha_k}\beta_j^{n(\alpha_j-\alpha_k)}\Gamma(\alpha_k)^n}{\pi_k\alpha_k^{n\alpha_k}\Gamma(\alpha_j)^n}+n\hat\mu_{j,k}\right)\right),
\label{eq:Pcf}
\end{empheq}
where $\hat\mu_{j,k}$ and $\hat\sigma_{j,k}^2$ are the mean and variance of $\hat y_i^{(j,k)}=\alpha_{j,k}\log x_i-\left(\frac{\alpha_j-\alpha_k}{\alpha_j}\right)\beta_j x_i$, respectively, which can also be derived analytically using the scheme given in Appendix~\ref{sec:pdf}.

\subsection{Estimate-then-Classify: Using MSPRTC}
\label{sec:ETC_MSPRT}

For the MSPRTC, we need to update the estimated posterior probabilities after collecting each new PU traffic period if all the estimated posterior probabilities defined as
\begin{align}
\hat p_n^j\triangleq{\pi_j\prod\limits_{i=1}^{n}f_j(x_i|\mathbf{\hat\Theta}_j)}{\left[\sum\limits_{l=1}^{M}\pi_l\left(\prod\limits_{i=1}^{n}f_l(x_i|\mathbf{\hat\Theta}_l)\right)\right]^{-1}},
\label{eq:pos_est}
\end{align}
are less than or equal to the threshold. The complete algorithm for ETC-based MSPRTC is listed in Algorithm~\ref{alg:1}. 
\begin{algorithm}[t]
\caption{ETC-based MSPRTC}
\label{alg:1}
\begin{algorithmic}[1]
\small
\Procedure{Classifier}{$\mathbf{x},f_j(x|\mathbf{\Theta}_j),M,\mathbf{\Omega},\gamma$}
\State $i\gets 1$
\State $\mathbf{x}\gets x_1$
\State {\bf Calculate} $\mathbf{\hat\Theta}_j$ using (\ref{eq:MLE}), $\forall j\in\{1,2,\cdots,M\}$
\State {\bf Calculate} estimated posteriori probability $\hat p_i^j$ using (\ref{eq:pos_est})
\While{$\hat p_i^j\leq \frac{1}{1+\gamma}\forall j\in\{1,2,\cdots,M\}$}
\State $i\gets i+1$
\State $\mathbf{x}\gets(x_1,x_2,\cdots,x_i)^T$
\State {\bf Calculate} $\mathbf{\hat\Theta}_j$ using (\ref{eq:MLE}), $\forall j\in\{1,2,\cdots,M\}$
\State {\bf Calculate} estimated posteriori probability $\hat p_i^j$ using (\ref{eq:pos_est})
\EndWhile
\State $N_A\gets i$ \Comment Stopping time
\State $m\gets\arg\max\limits_j \hat p_{N_A}^j$
\State $\nu\gets\mathcal{H}_m$ \Comment Final decision
\EndProcedure
\end{algorithmic}
\end{algorithm}
\subsection{Average Likelihood Function: Traffic Classification with Prior Knowledge on Distribution of PU Traffic Parameters}
\label{sec:classification_prior}

We now consider a case when the PU traffic parameters $\beta_j$ are no longer constants, but instead follow a certain distribution. When the distribution of the PU traffic parameter is known, such knowledge can be exploited by averaging the conditional likelihood function with respect to the distribution of the PU traffic parameter, which can better describe the behavior for each hypothesis. The proposed ALF under $\mathcal{H}_j$ is defined as
\begin{align}
h_j(x)\triangleq\int_{-\infty}^{\infty}f_j(x|\mathbf{\Theta}_j)q_j(\beta_j)d\beta_j,
\label{eq:ALF}
\end{align}
where $q_j(\beta_j)$ is the PDF for $\beta_j$. Hence the likelihood function in~(\ref{eq:LF}) for MLC and the posterior probability in (\ref{eq:pos}) for MSPRTC are modified by replacing likelihood function $f_j(x|\mathbf{\Theta}_j)$ with ALF $h_j(x)$. As an example, assuming $\beta_j\sim\mathcal{U}(L_j,U_j)$ then~(\ref{eq:ALF}) can be derived using~(\ref{eq:lemma_integral}) as
\begin{align}
h_j(x)&=\int_{L_j}^{U_j}f_j(x|\mathbf{\Theta}_j)q_j(\beta_j)d\beta_j
\nonumber\\&=\frac{x^{\alpha_j-1}}{(U_j-L_j)\Gamma(\alpha_j)}\int_{L_j}^{U_j}\beta_j^{\alpha_j}e^{-\beta_j x}d\beta_j\notag\\
&=\frac{\Gamma(\alpha_j+1,L_j x)-\Gamma(\alpha_j+1,U_j x)}{(U_j-L_j)\Gamma(\alpha_j)x^2}.
\label{eq:MSHD}
\end{align}
Note that the average SH distance with ALF can be calculated by using (\ref{eq:MSHD}) to replace $f_j(x|\mathbf{\Theta}_j)$ in (\ref{eq:SHD}). Also note that for the average SH distance with ALF we were unable to find a closed-form expression and it can only be computed through numerical integration.

\section{Numerical Results}
\label{sec:numerical_results}

We now present MATLAB-based numerical results for the performance of the proposed PU traffic classification algorithms. We assume $M=3$, as in~\cite[Sec. VIII]{Veeravalli94} in which two distributions are considered as special cases, that is where: (i) $\alpha_1=1$, i.e., exponential distribution, and (ii) $\alpha_2=2$, i.e., Erlang distribution. Furthermore, we design two test scenarios for the classifiers, i.e., Test I and Test II, with a relatively large and small average distribution distance among hypotheses, respectively. The average distance among hypotheses is evaluated through average SH distance, $H^2$, which is calculated in Appendix~\ref{sec:SH_2_Gamma}. The PU traffic parameters for Test I and Test II are summarized in Table~\ref{table1} for the PU traffic with stable parameters and in Table~\ref{table2} for PU traffic with fluctuating parameters, respectively. The unit for $\beta_i$ is second$^{-1}$. We assume that each hypothesis has the same prior probability, i.e., $\pi_j=\frac{1}{M}$, i.e. a maximum entropy case. Observe that for Test II, all hypotheses have the same first moment in order to have a small average distance among hypotheses, which is different from Test I. In our simulations, PU traffic periods are generated randomly from three distributions in one realization. In case of PU sampled process we generate it by adding two uniformly distributed random variables at the beginning and the end of PU traffic process, following strictly the simplifying assumption from Section~\ref{sec:noise_modeling}. Each simulation point is obtained by method of batch means (unless otherwise stated) averaging 50 classification runs, each having at least 2000 realizations for a confidence interval of 0.1. 

\subsection{Results Reproducibility and Open Code Access}
\label{sec:result_reproducibility}

In addition, for the reproducibility of results, the source code used in generating all figures is (i) available upon request or (ii) via this ArXiv submission. The code allows the reader to generate results for a desired set of variables and experiment with the implementation and the accuracy of the developed classifiers. Any future corrections and updates to the source code and the paper will be also available therein.
\begin{table}
\caption{Traffic Parameters (Stable)}
\centering
\begin{tabular}{c | c | c}
\hline
Function &Parameters (Test I) & Parameters (Test II)\\
\hline\hline
exponential & $\alpha_1=1$, $\beta_1=0.4$ & $\alpha_1=1$, $\beta_1=0.4$\\
Erlang & $\alpha_2=2$, $\beta_2=0.3$ & $\alpha_2=2$, $\beta_2=0.8$\\
gamma & $\alpha_3=0.8$, $\beta_3=0.5$ & $\alpha_3=0.5$, $\beta_3=0.2$\\ 
\hline
Average $H^2$ & $0.1799$ & $0.0695$\\
\hline
\end{tabular}
\label{table1}
\end{table}

\begin{table}
\caption{Traffic Parameters (Fluctuating)}
\centering
\begin{tabular}{c | c | c}
\hline
Function &Parameters (Test I) & Parameters (Test II)\\
\hline\hline
exponential & $\alpha_1=1, \beta_1\sim\mathcal{U}(0.4,0.9)$ & $\alpha_1=1, \beta_1\sim\mathcal{U}(0.4,0.9)$\\
Erlang & $\alpha_2=2$, $\beta_2\sim\mathcal{U}(0.1,0.3)$ & $\alpha_2=2$, $\beta_2\sim\mathcal{U}(1.2,1.4)$\\
gamma & $\alpha_3=0.2$, $\beta_3\sim\mathcal{U}(0.2,0.5)$ & $\alpha_3=3$, $\beta_3\sim\mathcal{U}(1.1,2.8)$\\
\hline
Average $H^2$ & $0.4482$ & $0.0379$\\
\hline
\end{tabular}
\label{table2}
\end{table}

\subsection{Traffic Classification Performance with Perfectly Sampled PU Traffic Periods and Parameters}

\begin{figure}
\centering
\subfigure[Test I]{\includegraphics[width=0.75\columnwidth]{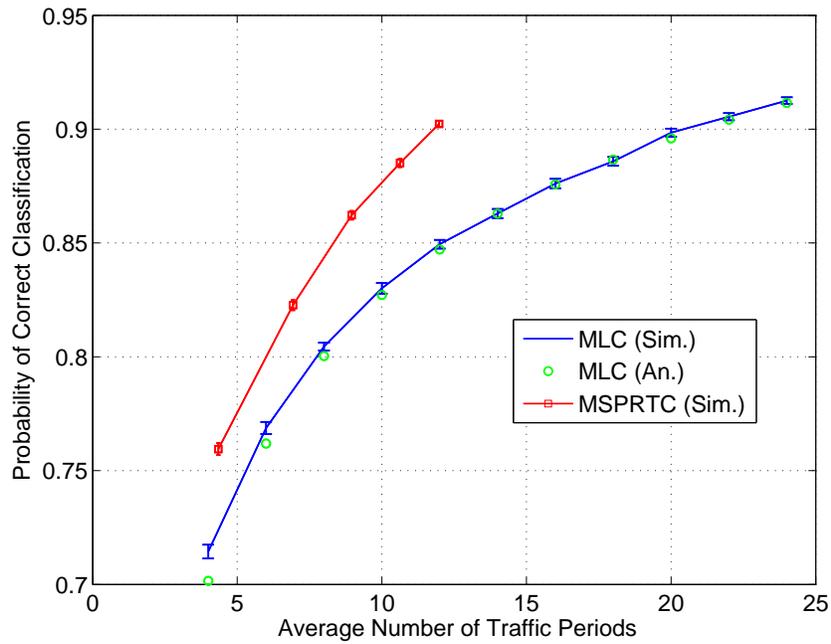}\label{fig:fig1a}}\vspace{-1em}
\subfigure[Test II]{\includegraphics[width=0.75\columnwidth]{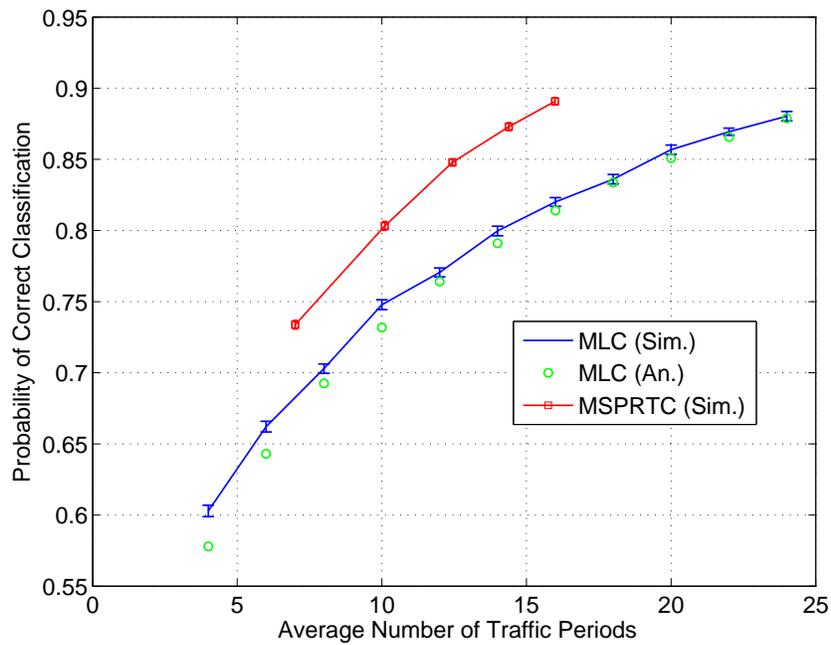}\label{fig:fig1b}}\vspace{-0.5em}
\caption{Probability of correct classification with the average number of PU traffic periods, under perfect knowledge of PU traffic periods and parameters. MLC is compared with MSPRTC. PU traffic parameters used in simulations are presented in Table~\ref{table1}. Simulation results (Sim.) are plotted to verify analytical results (An.).}
\label{fig:Perfect}
\end{figure}

In Fig.~\ref{fig:Perfect} we present the classification performance under perfect knowledge of PU traffic parameters and perfect sampling of traffic ON/OFF periods as a function of traffic periods $n$. First, we observe that under both tests the simulated MLC performance matches our derived analytical performance. Second, the MSPRTC performs better than MLC since it can achieve the same $P_c$ using less number of PU traffic periods. Finally, our results prove the intuitive observation that for a smaller average distance among hypotheses, shown in Fig.~\ref{fig:fig1b}, a higher number of PU traffic periods is needed to classify the correct hypotheses\footnote{Note that in the MATLAB implementation we are constrained by the numerical precision of 32 bit unsigned integers (due to frequent exponentiations of very small numbers) thus the analytical results are not realizable for large values of $n$. Also, note that while plotting the analytical results for the MLC classifier, we have used a simulation to generate statistics for mean and variance for $y_{i}^{(j,k)}$ and $\tilde y_{i}^{(j,k)}$, to speed up figure generation. More details are provided in the code accompanying this paper.}.

\subsection{Traffic Classification Performance with PU Traffic Period Estimation and Perfect Knowledge of Parameters}

\begin{figure}
\centering
\subfigure[Test I]{\includegraphics[width=0.75\columnwidth]{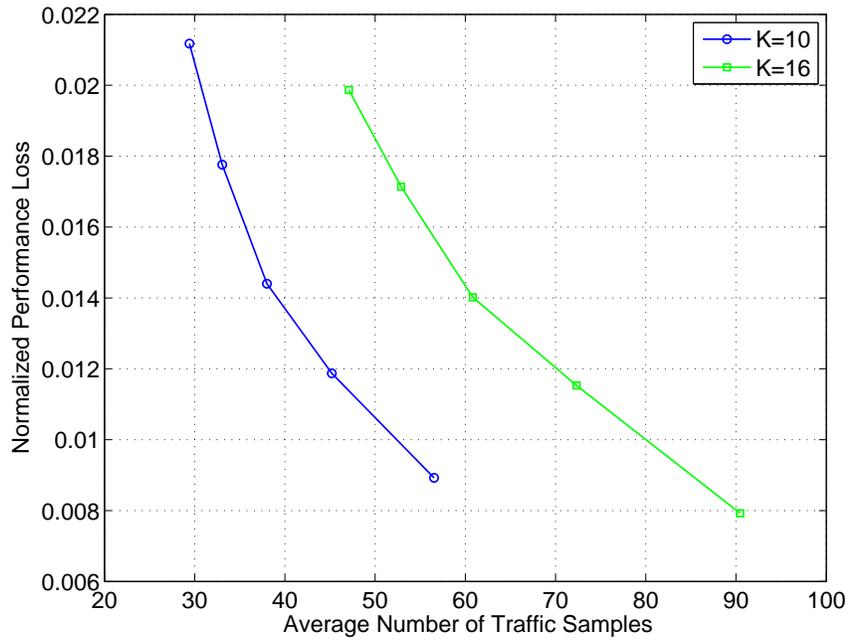}\label{fig:fig2a}}\vspace{-1em}
\subfigure[Test II]{\includegraphics[width=0.75\columnwidth]{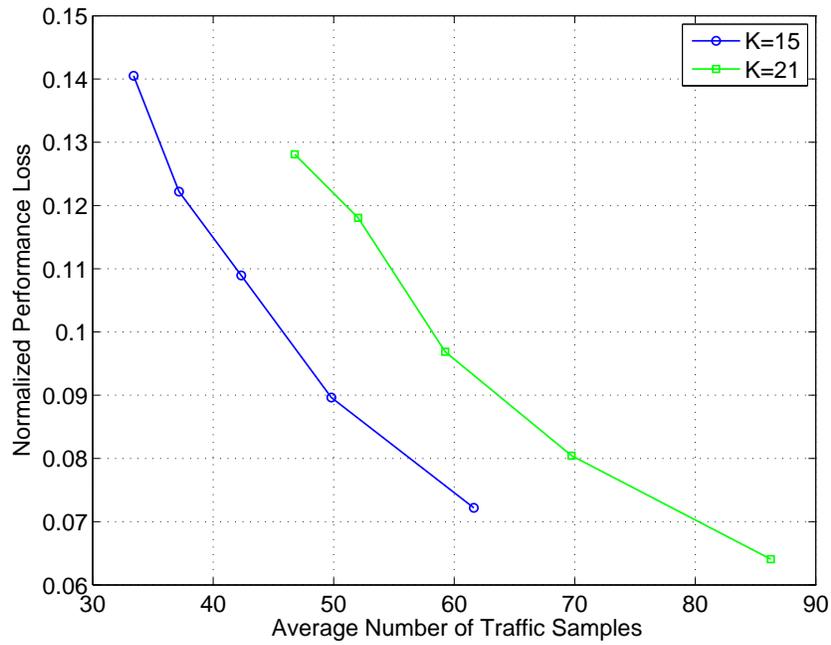}\label{fig:fig2b}}\vspace{-0.5em}
\caption{Normalized classification performance loss with average number of traffic samples using the minimum variance estimator for the MLC under perfect knowledge of PU traffic parameters. The PU traffic parameters used are given in Table~\ref{table1}. All results were obtained by simulations.}
\label{fig:loss}
\end{figure}

Fig.~\ref{fig:loss} shows the normalized performance loss $L\triangleq\frac{P_c-\tilde P_c}{P_c}$ for MLC with the average number of traffic samples\footnote{In this case we do not plot the confidence intervals as we plot the difference between the two means.} $\mathbb{E}\{N\}$, which are both functions of $T_s$. We consider two cases of PU traffic periods: (i) $K=10$ and (ii) $K=16$. First, as the average number of PU traffic samplings increases, which means we adopt a small sampling period $T_s$, $L$ decreases. This is because we have higher resolution for sampling to estimate the PU traffic periods, thus resulting in a more accurate classification. Second, we observe that the performance for higher number of PU traffic periods is more sensitive to the PU traffic period estimation error. Therefore more PU traffic samples for higher number of PU traffic periods are needed to achieve the same performance as with a lower number of PU traffic periods. Finally, we show that for a small average distance among hypotheses, the performance loss is large since in this case the PU traffic classification is more sensitive to the period estimation errors. 

\begin{figure}
\centering
\subfigure[Test I]{\includegraphics[width=0.75\columnwidth]{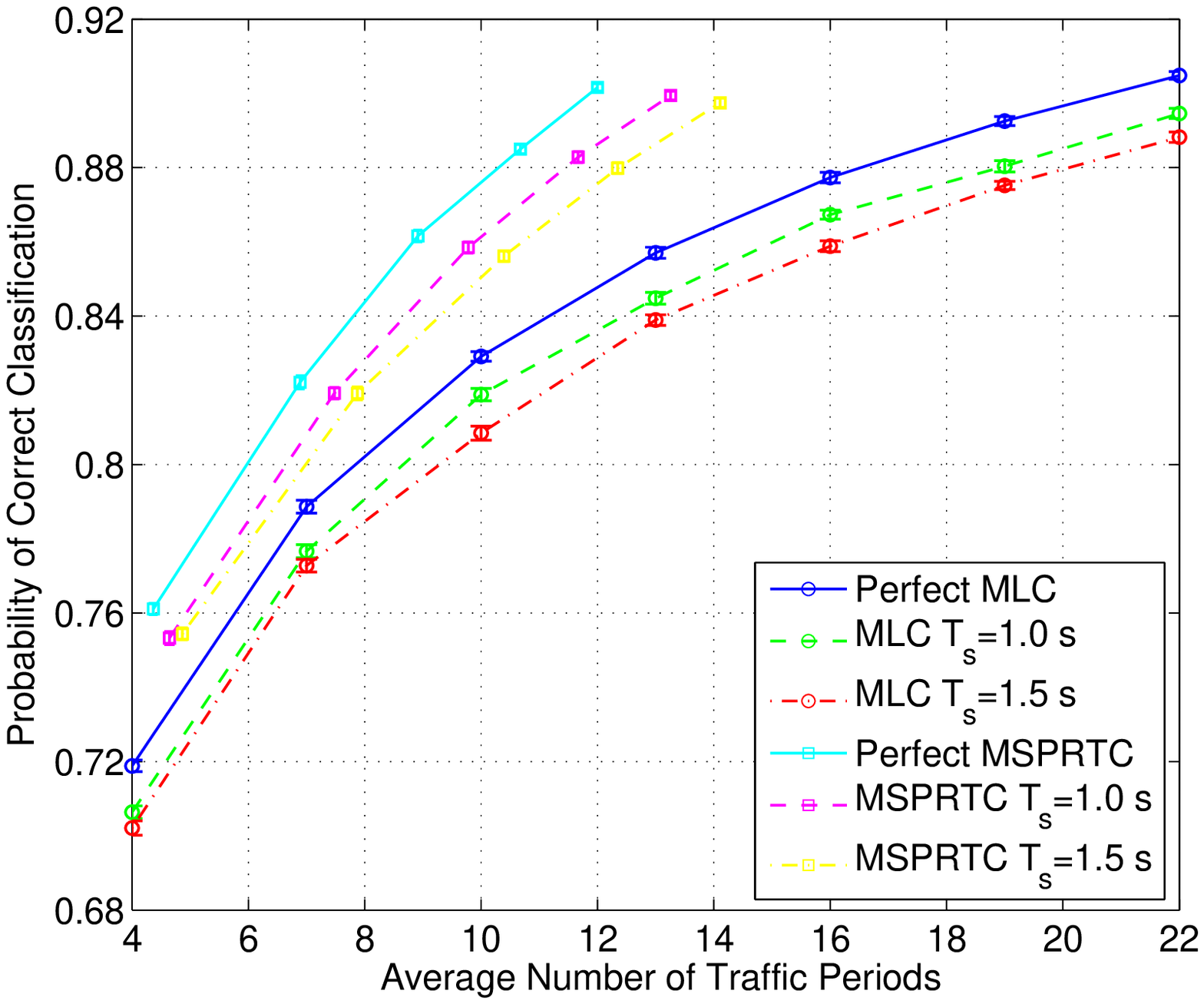}\label{fig:fig3a}}\vspace{-1em}
\subfigure[Test II]{\includegraphics[width=0.75\columnwidth]{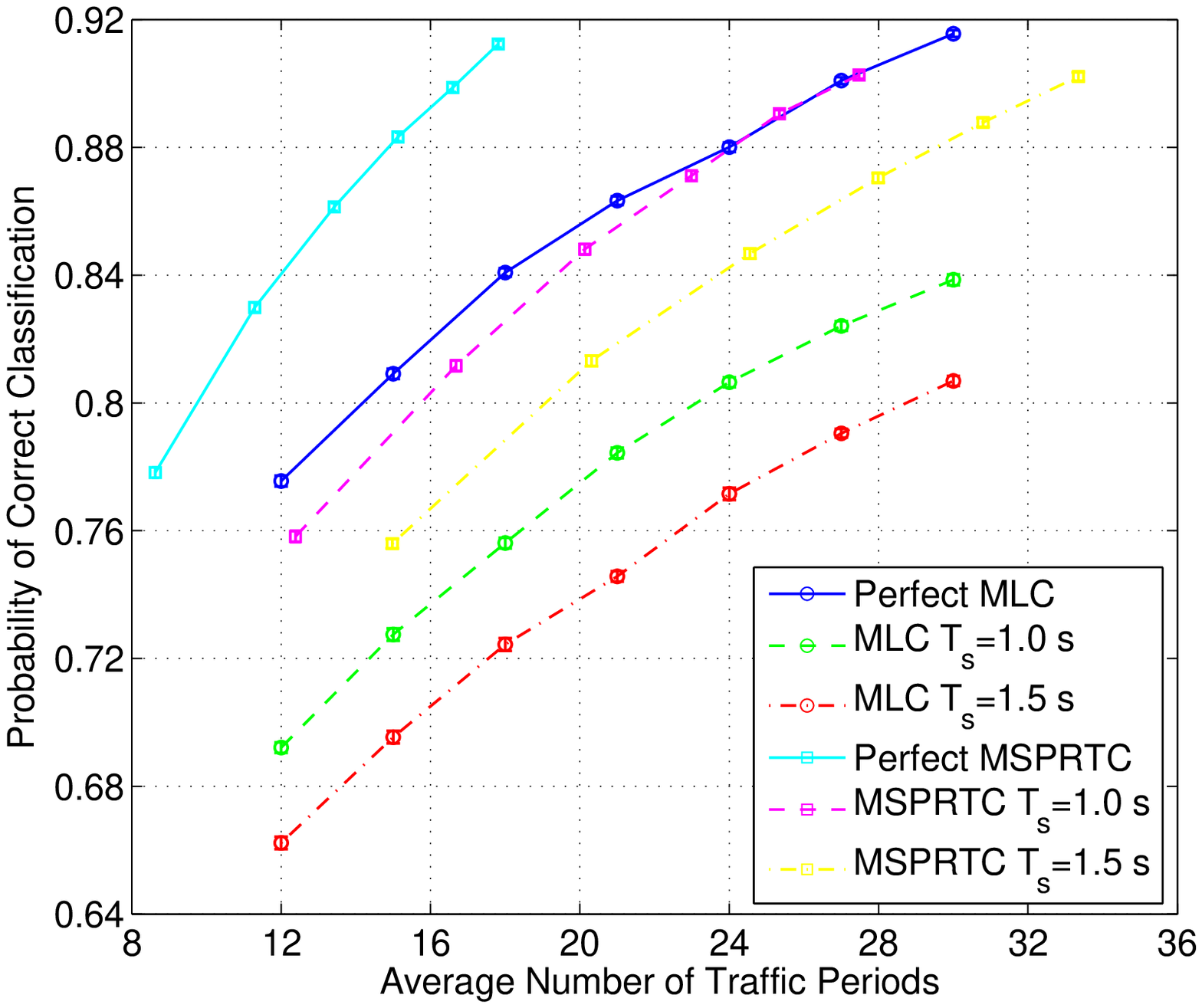}\label{fig:fig3b}}\vspace{-0.5em}
\caption{Probability of correct classification with the average number of PU traffic periods using the minimum variance estimator, under perfect knowledge of PU traffic parameters. MLC is compared with MSPRTC. The PU traffic parameters used are given in Table~\ref{table1}. All results were obtained by simulations.}
\label{fig:Pc_period}
\end{figure}

In Fig.~\ref{fig:Pc_period} we compare MLC and MSPRTC under sampling. First, as the sampling period increases, the performance of both classifiers decreases. Naturally, a longer sampling period will result in a larger estimation noise. Second, we observe that MSPRTC is as sensitive as MLC to the period estimation error. This is because both MSPRTC and MLC adopt the same likelihood function for classification, which requires accurate knowledge of the true distributions. If the noise is added into the observation, it will distort the original PDF even worse when the distance among hypotheses is small.

\subsection{A Design Guideline Example for Traffic Classification using ML Classifier}
\label{sec:design_guideline_traffic_classification}

\begin{figure}
\centering
\subfigure[Sensing constraint on observation time]{\includegraphics[width=0.75\columnwidth]{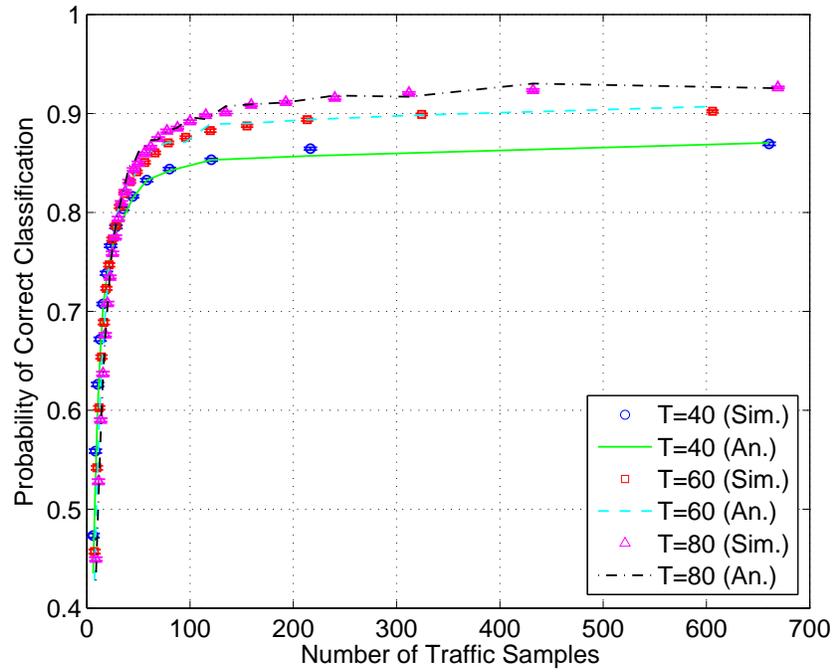}\label{fig:fig4a}}\vspace{-1em}
\subfigure[Sensing constraint on number of traffic samples]{\includegraphics[width=0.75\columnwidth]{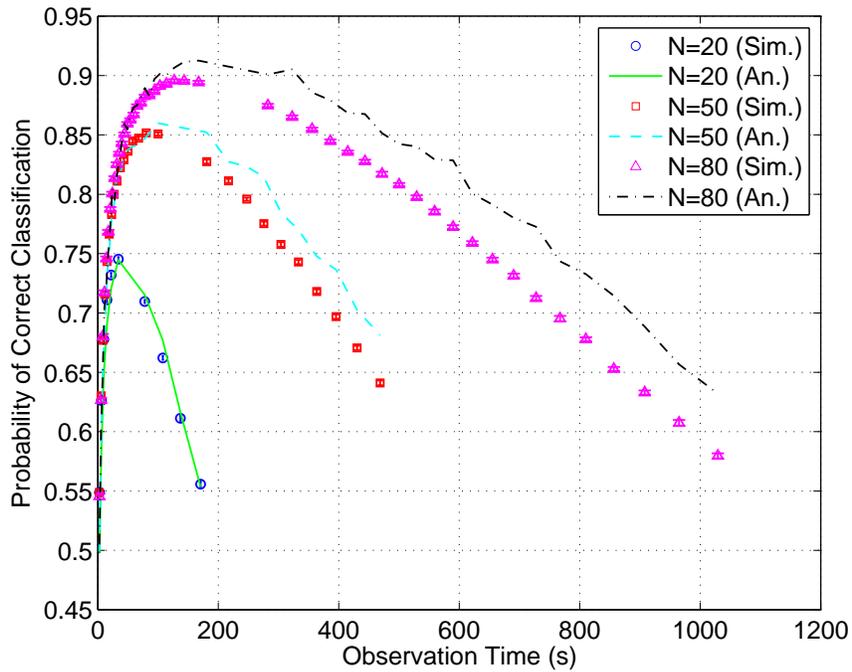}\label{fig:fig4b}}\vspace{-0.5em}
\caption{Probability of correct classification with number of traffic samples in Fig.~\ref{fig:fig4a} and observation time in Fig.~\ref{fig:fig4b} with perfect knowledge of traffic parameters using MLC. The PU traffic parameters used are shown as Test I in Table~\ref{table1}. Simulation results (Sim.) are plotted to verify analytical results (An.).}
\label{fig:guideline}
\end{figure}

We provide two examples for the design guideline shown in Section~\ref{sec:Guideline}. First we consider the case where, given observation time $T$, we need to find the number of traffic samples and therefore sampling period $T_s$, to achieve a certain probability of correct classification. In Fig.~\ref{fig:fig4a} we observe that as the number of traffic samples increases the classification performance improves. This is because as the number of traffic samples increases, the period estimation errors decrease, and at the same time, we can obtain more PU traffic periods as the PU traffic period mis-detection rate decreases which is shown in~(\ref{eq:K}). Furthermore, as the observation time increases, the classification performance also increase. Although in this case the estimation error increases, the obtained traffic periods increases. This is because the latter factor has more influence on the classification performance. In this traffic scenario, for example, given the timing constraint $T=60$\,seconds we need at least $N=350$ traffic samples to achieve the performance $\epsilon=0.90$. This means the constraint for the sampling rate $T_s$ to sample this traffic should be no less than $\frac{60}{350-1}=0.1719$\,seconds to achieve the classification performance of $\epsilon=0.90$.

Second we consider the case where, given the number of samples, we need to find the observation time to achieve a certain classification performance. From Fig.~\ref{fig:fig4b}, the performance is a concave curve with respect to the observation time. This can be explained by the behavior of~(\ref{eq:K}). In~(\ref{eq:K}), $\mathbb{E}\{K\}$ versus $T$ has a similar shape as $\tilde P_c$ versus $T$. However, to figure out the classification performance, not only $\mathbb{E}\{K\}$ but also the sampling period $T_s$ needs to be considered to determine the classification performance. Initially, as $T$ increases, $\mathbb{E}\{K\}$ increases, and $T_s$ increases. Since the effect of $\mathbb{E}\{K\}$ is more significant, the classification performance increases. As $T$ increases through the maximum point of $\mathbb{E}\{K\}$, $\mathbb{E}\{K\}$ starts to decrease. In this region $T_s$ also increases. Therefore the performance will decrease since we obtain less traffic periods with higher estimation errors. In this traffic scenario, for example, given the energy constraint $N=50$, we can solve for the optimal observation time $T=100$\,seconds to achieve the maximal performance $\epsilon=0.86$. This means the optimal sampling rate $T_s$ to sample this traffic should be set as $\frac{100}{50-1}=2.04$\,seconds to achieve $\epsilon=0.86$. Larger and smaller $T_s$ than the optimal $T_s$ will both degrade the classification performance.

Finally, we see that our proposed analytical approximation matches the simulation results for small values of $T_s$. But as $T_s$ increases, shown in Fig.~\ref{fig:fig4b}, the analytical results start to deviate from the simulation results, refer again to Section~\ref{sec:MLC_Error}.

\subsection{Traffic Classification with Perfect PU Periods and No Knowledge of Parameters}
\label{sec:classification_performance_blind_parameters}

\begin{figure}
\centering
\subfigure[Test I]{\includegraphics[width=0.75\columnwidth]{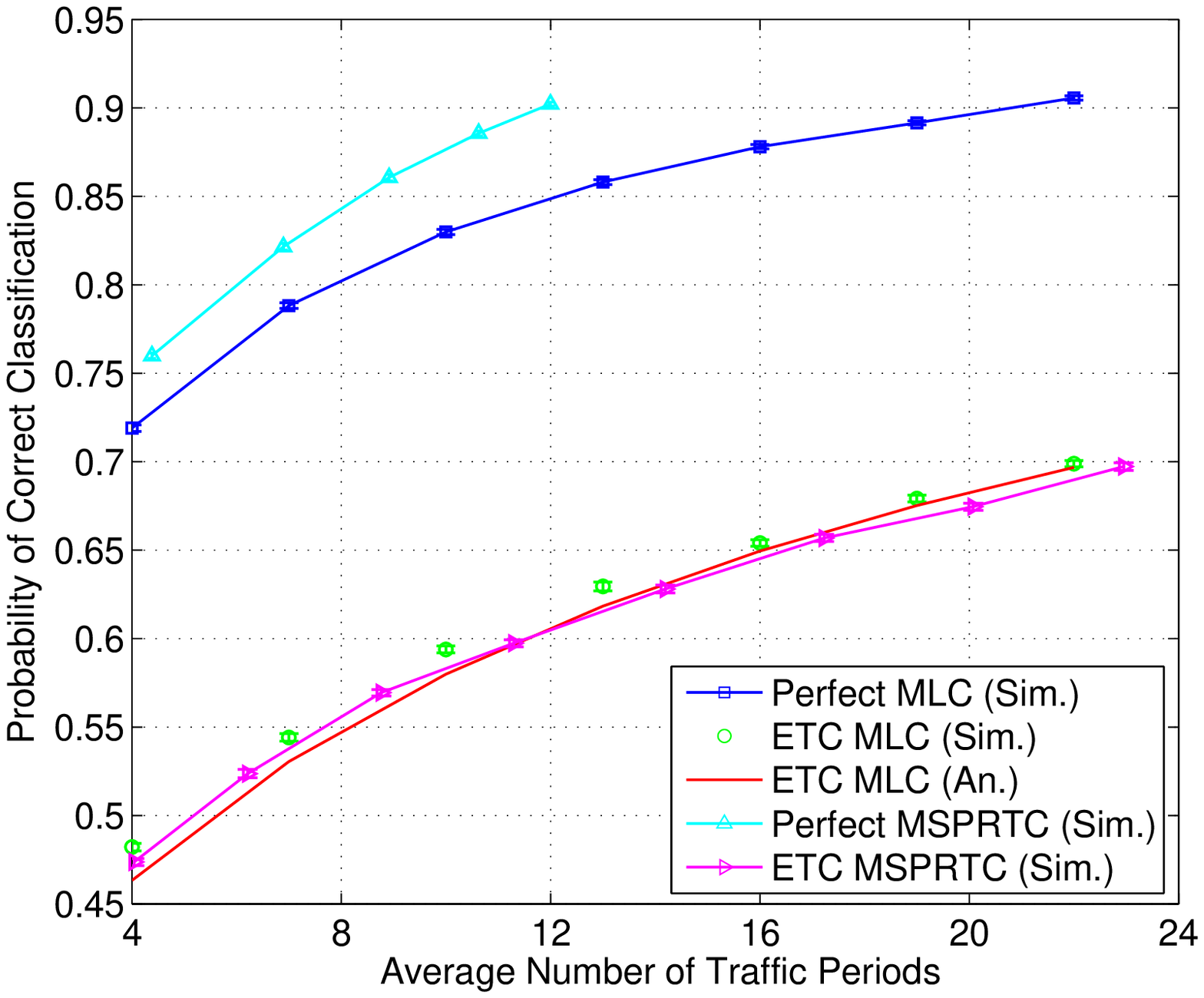}\label{fig:fig5a}}\vspace{-1em}
\subfigure[Test II]{\includegraphics[width=0.75\columnwidth]{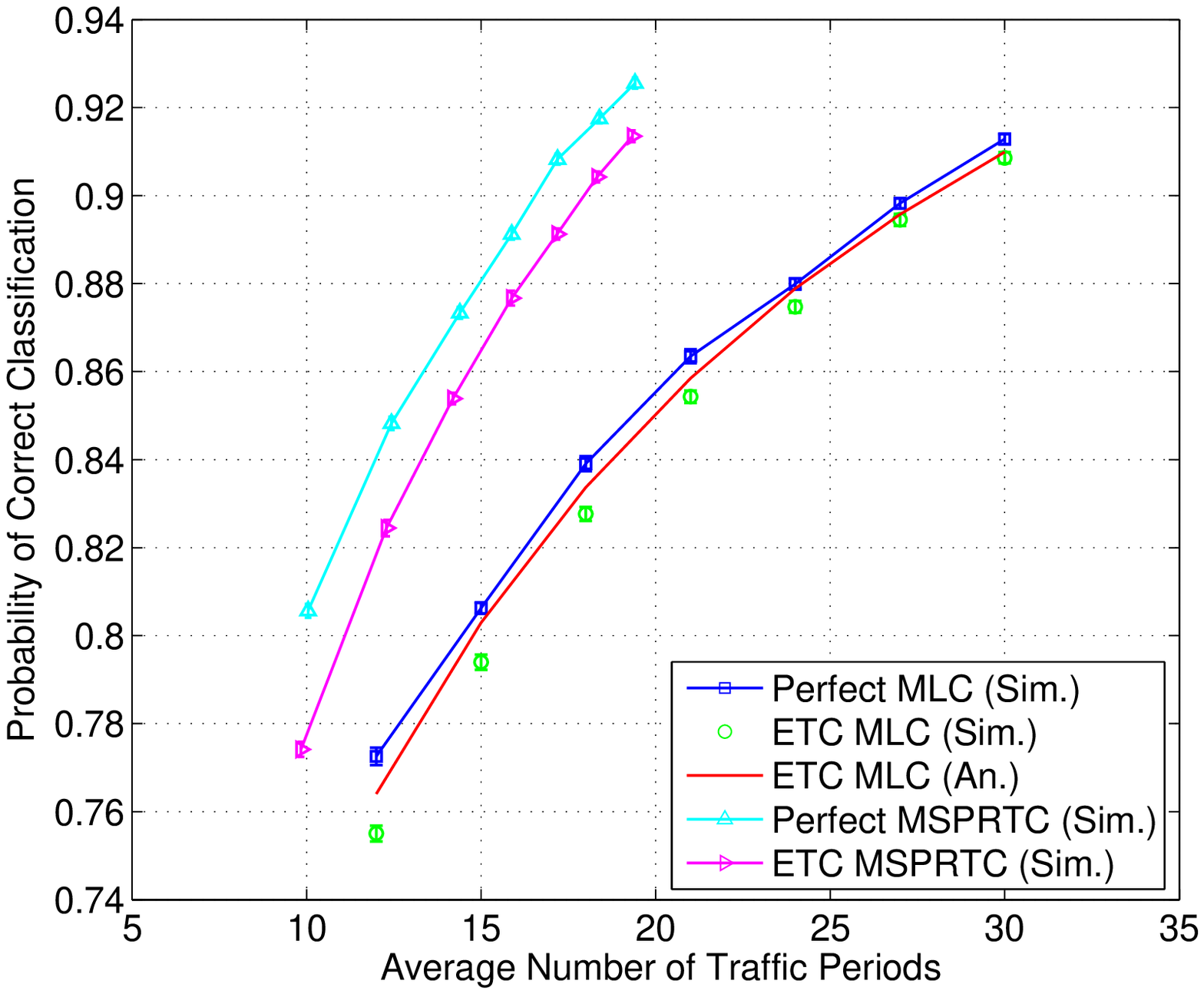}\label{fig:fig5b}}\vspace{-0.5em}
\caption{Probability of correct classification with the average number of PU traffic periods, under no knowledge of PU traffic parameters $\beta_j$. MLC is compared with MSPRTC using ETC scheme. The PU traffic parameters used are shown in Table~\ref{table1}. Simulation results (Sim.) are plotted to verify analytical results (An.).}
\label{fig:blind}
\end{figure}

Fig.~\ref{fig:blind} presents the probability of correct classification with the average number of PU traffic periods assuming no knowledge of PU traffic parameters $\beta_j$. We compare MLC and MSPRTC with perfect knowledge of PU traffic parameters and the ETC method with no knowledge of traffic parameters $\beta_j$. First, we note that ETC-based method performs worse than methods using perfect parameters. Second, the ETC-based MSPRTC outperforms MLC as the distance among hypotheses is small, otherwise they perform similarly. Third, the simulation results for ETC-based MLC matches our proposed analytical results in~(\ref{eq:Pcf}), since the number of PU traffic periods is large enough for parameter estimation. Finally, we can observe that ETC-based method will perform worse under Test I than Test II, compared with the perfect classifiers. This is because in Test II the first moments for all hypotheses are set to be the same, hence the estimated parameters will be close to the true parameters for all hypotheses. But this is not the case for Test I since the first moments are more different for all hypotheses---which means a small parameter estimation error will cause a large classification performance degradation.

\subsection{Traffic Classification Performance with Perfect PU Traffic Periods and Prior Knowledge of Traffic Parameters}
\label{sec:sec:classification_performance_perfect_periods}

\begin{figure}
\centering
\subfigure[Test I]{\includegraphics[width=0.75\columnwidth]{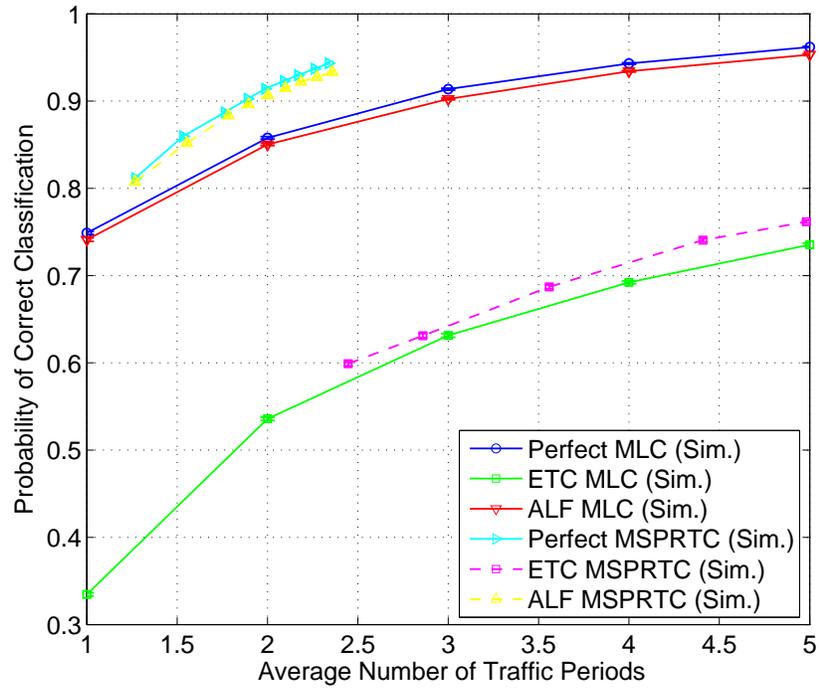}\label{fig:fig6a}}\vspace{-1em}
\subfigure[Test II]{\includegraphics[width=0.75\columnwidth]{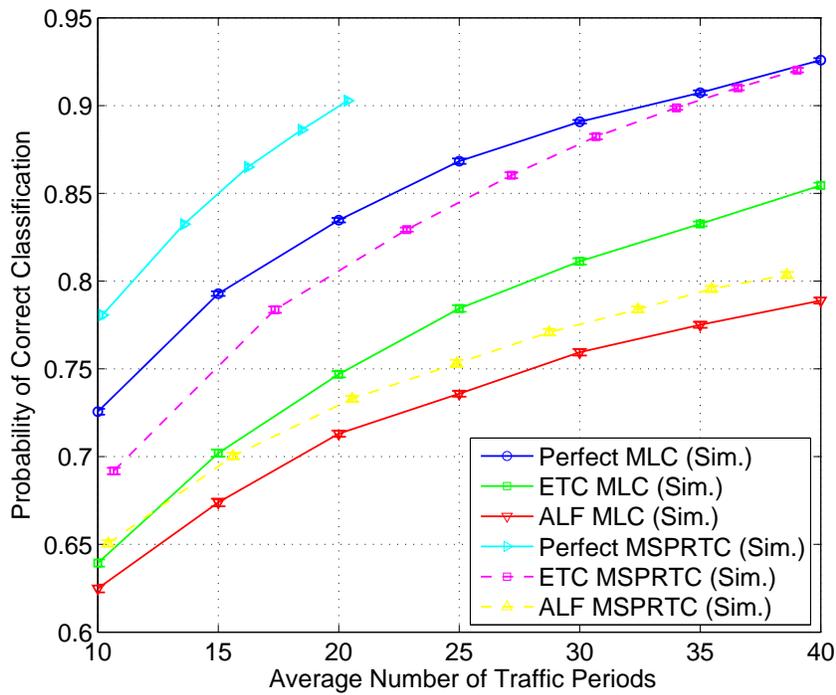}\label{fig:fig6b}}\vspace{-0.5em}
\caption{Probability of correct classification with the average number of PU traffic periods, with prior knowledge of PU traffic parameters $\beta_j$. MLC is compared with MSPRTC using ETC and ALF schemes. The PU traffic parameters used are shown in Table~\ref{table2}.}
\label{fig:Partial}
\end{figure}

In Fig.~\ref{fig:Partial} we present the classification performance comparisons assuming prior knowledge about the distribution of PU traffic parameters $\beta_j$. We note that ALF-based classifiers are better than ETC-based classifiers under Test I, and the result is opposite under Test II. This is because of the fact that ALF can capture most PU traffic parameter information if the distance among hypotheses is large, i.e., the Test I case. If the distance among hypotheses is small, as in Test II, ETC-based method provides a more accurate PU traffic parameter estimation. 

\section{Conclusions}
\label{sec:conclusions}

We propose novel primary user (PU) traffic classification algorithms which are based on the maximum likelihood function and multi-hypothesis sequential probability ratio test classifiers, and we consider cases where the PU traffic periods and PU traffic parameters need to be estimated. In addition, we analyze a sampling technique to estimate PU traffic periods, and a minimum variance period estimator is derived to design a traffic classifier given sensing constraints such as the number of traffic samples or observation time. Furthermore, we propose two classifiers, estimate-then-classify (ETC) and average likelihood function (ALF) classifiers to handle the cases when there is only no/partial knowledge of PU traffic parameters.

To conclude, for PU traffic with constant and known parameters, MSPRTC, a more complicated classifier than MLC is recommended in terms of classification performance both with and without period estimation. For PU traffic with prior knowledge of parameters, the ALF-based classifier is suitable for traffic classification when the average distance among hypotheses is large. If the average distance among hypotheses is small, the ETC-based classifier is preferred to provide a good classification performance. 

\appendices

\section{Derivation of Mean and Variance for the Distribution of $y_i^{(j,k)}$}
\label{sec:pdf}

To derive the mean and variance for $y_i^{(j,k)}$, we need to derive its PDF first. Here we ignore the index $i$ for convenience since all $y_i^{(j,k)}$ have the same distribution. Since we know the PDF for $x$ under hypothesis $\mathcal{H}_j$, we can apply the change of variable technique to derive the PDF for $y^{(j,k)}$ as 
\begin{align}
f_{Y^{(j,k)}}(y^{(j,k)})&=\left|\frac{\partial}{\partial y^{(j,k)}}h^{-1}\left(y^{(j,k)}\right)\right|\nonumber\\&\quad\times f_j\left(h^{-1}\left(y^{(j,k)}\right)|\mathbf{\Theta}_j\right),
\label{eq:yPDF}
\end{align}
where $|\cdot|$ is the absolute value function, $h(x)=\alpha_{j,k}\log x-\beta_{j,k}x$, and $h^{-1}$ is the inverse function of $h$. To find $h^{-1}$, we introduce first the following Lemma.
\begin{lemma}
The inverse function for $h(x)=\alpha\log(x)-\beta x$, $\forall \alpha\neq 0,\beta\neq0, x>0$, is (i) when $\frac{\alpha}{\beta}<0$, $h^{-1}(y)=-\frac{\alpha}{\beta}W\left(0,e^{\frac{y}{\alpha}+\log\left(\frac{-\beta}{\alpha}\right)}\right)$, and (ii) when $\frac{\alpha}{\beta}>0$ $h^{-1}(y)=-\frac{\alpha}{\beta}W\left(0,e^{\frac{y}{\alpha}+\log\left(\frac{-\beta}{\alpha}\right)}\right)$, if $h^{-1}(y)\leq\frac{\alpha}{\beta}$, and $h^{-1}(y)=-\frac{\alpha}{\beta}W\left(-1,e^{\frac{y}{\alpha}+\log\left(\frac{-\beta}{\alpha}\right)}\right)$, otherwise, where $W(k,y)$ is a Lambert W function of branch $k$, where $k$ is an integer for complex $y$ and $k\in\{0,-1\}$ for real $y$ (refer to MATLAB's \texttt{lambertw} function implementation description)~\cite[Eq. (1.5)]{Knuth96}.
\label{lamme_inverse}
\end{lemma}
\begin{IEEEproof}
Consider the Wright omega function, $\omega(y)$~\cite[Eq. (1)]{corless_techrep}, which is defined as the unique solution to $y=\log(x) + x$, which can be also written recursively as
\begin{align}
y=\log(\omega(y))+\omega(y),
\label{eq:omega}
\end{align}
where $W(0, e^y)=\omega(y)$. Embedding $x=-\frac{\alpha}{\beta}\omega\left(\frac{y}{\alpha}+\log\left(\frac{-\beta}{\alpha}\right)\right)$ to the expression $\alpha\log(x)-\beta x$ we can show that
\begin{align}
&\alpha\log\left(-\frac{\alpha}{\beta}\omega\left(\frac{y}{\alpha}+\log\left(\frac{-\beta}{\alpha}\right)\right)\right)\nonumber\\&\quad-\beta\left(-\frac{\alpha}{\beta}\omega\left(\frac{y}{\alpha}+\log\left(\frac{-\beta}{\alpha}\right)\right)\right)\\
&\quad=\alpha\left(\log\omega\left(\frac{y}{\alpha}+\log\left(\frac{-\beta}{\alpha}\right)\right)+\omega\left(\frac{y}{\alpha}+\log\left(\frac{-\beta}{\alpha}\right)\right)\right)\nonumber\\&\qquad+\alpha\log\left(\frac{-\alpha}{\beta}\right)
\label{eq:simplify1}\\
&\quad=\alpha\left(\frac{y}{\alpha}+\log\left(\frac{-\beta}{\alpha}\right)\right)+\alpha\log\left(\frac{-\alpha}{\beta}\right)=y,\label{eq:simplify2}
\end{align}
where~(\ref{eq:simplify2}) stems directly from~(\ref{eq:omega}). Therefore we know $x$ is an inverse function.

Now, note that the function $h(x)$ is a concave function as $\alpha\geq 0$, and convex otherwise. Therefore, for $\alpha\geq 0$, there are two possible real-value solutions for $h(x)=y$: (i) one is located on the left hand side of the peak value for $h(x)$, i.e., $x=\frac{\alpha}{\beta}$, and (ii) another located on its right hand side. By definition of a Lambert W function, these two solutions are shown to be located on $k=0$ and $k=-1$ branches. For $\alpha<0$, there is only one solution on $k=0$ branch since $h(x)$ is a decreasing function. Note also that domain of y is (i) $[-\infty,a\log(a/b)-a]$ for $a,b>0$, (ii) $[a\log(a/b)-a,\infty]$ for $a,b<0$, and (iii) $[-\infty,\infty]$ otherwise.
\end{IEEEproof}

By applying the derivative of the Lambert W function, i.e., $\frac{\partial W(k,s)}{\partial s}=\frac{W(k,s)}{s(1+W(k,s))}$, and Lemma~\ref{lamme_inverse} to~(\ref{eq:yPDF}), we can derive the PDF for $y^{(j,k)}$ as
\begin{align}
f_{Y^{(j,k)}}(y^{(j,k)})&=\left|\frac{W\left(0,e^{B^{(j,k)}}\right)}{\beta_{j,k}\left(1+W\left(0,e^{B^{(j,k)}}\right)\right)}\right|\nonumber\\&\times f_j\left(-\frac{\alpha_{j,k}}{\beta_{j,k}}W\left(0,e^{B^{(j,k)}}\right)|\mathbf{\Theta}_j\right)\notag\\
&+I\left(\frac{\alpha_{j,k}}{\beta_{j,k}}\right)\left|\frac{W\left(-1,e^{B^{(j,k)}}\right)}{\beta_{j,k}\left(1+W\left(-1,e^{B^{(j,k)}}\right)\right)}\right|\nonumber\\&\times f_j\left(-\frac{\alpha_{j,k}}{\beta_{j,k}}W\left(-1,e^{B^{(j,k)}}\right)|\mathbf{\Theta}_j\right),
\label{eq:yfPDF}
\end{align}
where $B^{(j,k)}\triangleq\frac{y^{(j,k)}}{\alpha_{j,k}}+\log\left(\frac{-\beta_{j,k}}{\alpha_{j,k}}\right)$ (defined for presentation compactness), and $I(a)=1$ if $a\geq 0$ and $I(a)=0$ otherwise. 

We can finally derive the mean and variance using~(\ref{eq:yfPDF}) as
\begin{align}
\mu_{j,k}=\int_{-\infty}^{\infty}y^{(j,k)}f_{Y^{(j,k)}}\left(y^{(j,k)}\right)dy^{(j,k)},
\end{align}
\begin{align}
\sigma_{j,k}^2&=\int_{-\infty}^{\infty}\left(y^{(j,k)}\right)^2 f_{Y^{(j,k)}}\left(y^{(j,k)}\right)dy^{(j,k)}\nonumber\\&\qquad-\left(\mu_{j,k}\right)^2,
\end{align}
respectively, through numerical integration.

\section{Derivation of Squared Hellinger Distance between Two Gamma Distributions}
\label{sec:SH_2_Gamma}

The SH distance for two probability distributions is defined as~\cite[Ch. 14.5, pp. 211]{vandervaart_cup_2002}
\begin{align}
H^2(f_j(x|\mathbf{\Theta}_j),&f_k(x|\mathbf{\Theta}_k))\nonumber\\&\triangleq1-\int_{-\infty}^{\infty}\sqrt{f_j(x|\mathbf{\Theta}_j)f_k(x|\mathbf{\Theta}_k)}dx,
\label{eq:SHD}
\end{align}
again, note that the 0.5 constant is omitted for convenience as remarked in~\cite[Ch. 3.3, pp. 61]{pollard_cup_2002}). Before calculating the closed-form expression of SH distance for two gamma distributions we introduce the following integral
\begin{align}
\int_{a}^{\infty}x^{\rho}e^{-\mu x}dx=\frac{\Gamma(\rho+1,a\mu)}{\mu^{\rho+1}},
\label{eq:lemma_integral}
\end{align}
where $\Gamma(\rho+1,x)=\int_{x}^{\infty}t^{\rho}e^{-t}dt$ is the incomplete gamma function. Integral~(\ref{eq:lemma_integral}) can be derived through calculating the incomplete gamma function by the change of variable technique.

From the definition of (\ref{eq:SHD}) the SH distance for two distributions $f_j(x|\mathbf{\Theta}_j)$ and $f_k(x|\mathbf{\Theta}_k)$ can be derived as
\begin{align}
H^2(&f_j(x|\mathbf{\Theta}_j),f_k(x|\mathbf{\Theta}_k))\nonumber\\&=1-C(\mathbf{\Theta}_j,\mathbf{\Theta}_k)\int_0^{\infty}x^{\frac{\alpha_j+\alpha_k}{2}-1}e^{-\left(\frac{\beta_j+\beta_k}{2}\right) x}dx,
\label{eq:SHDS}
\end{align}
where $C(\mathbf{\Theta}_j,\mathbf{\Theta}_k)=\sqrt{\frac{\beta_j^{\alpha_j}\beta_k^{\alpha_k}}{\Gamma(\alpha_j)\Gamma(\alpha_k)}}$. Applying~(\ref{eq:lemma_integral}) with $\rho=\frac{\alpha_j+\alpha_k}{2}-1$ and $\mu=\frac{\beta_j+\beta_k}{2}$ to~(\ref{eq:SHDS}) the SH distance in~(\ref{eq:SHDS}) can be simplified to
\begin{align}
H^2(f_j(x|\mathbf{\Theta}_j),&f_k(x|\mathbf{\Theta}_k))\nonumber\\&=1-C(\mathbf{\Theta}_j,\mathbf{\Theta}_k)\frac{\Gamma(\frac{\alpha_j+\alpha_k}{2})}{\left(\frac{\beta_j+\beta_k}{2}\right)^{\frac{\alpha_j+\alpha_k}{2}}}.
\end{align}
Note that the average SH distance with ALF, which is used to represent the average distance among hypotheses in Table~\ref{table2}, can be calculated by using (\ref{eq:MSHD}) to replace $f_j(x|\mathbf{\Theta}_j)$ in (\ref{eq:SHD}). Also note that the average SH distance with ALF has no closed-form expression and it can only be computed through numerical methods.

\section{Derivation of Expected Number of PU Traffic Samples under Sampling}
\label{sec:samples_sampling}

The expected average number of PU traffic samplings for one period $T_{\text{on}}$ under hypothesis $\mathcal{H}_j$ can be calculated as
\begin{align}
\mathbb{E}\{N|\mathcal{H}_j\}=\mathbb{E}\left\{\left\lfloor\frac{T_{\text{on}}}{T_s}\right\rfloor\right\}+1, 
\label{eq:ENHJ}
\end{align}
where $\lfloor\cdot\rfloor$ is the floor function. To calculate~(\ref{eq:ENHJ}) we first need to derive the following conditional probability, i.e.,
\begin{align}
\Pr&\left\{\left\lfloor\frac{T_{\text{on}}}{T_s}\right\rfloor=k|\mathcal{H}_j\right\}
=\Pr\left\{\frac{T_{\text{on}}}{T_s}-1<k\leq\frac{T_{\text{on}}}{T_s}|\mathcal{H}_j\right\}\nonumber\\&\qquad=\Pr\{kT_s\leq T_{\text{on}}<(k+1)T_s|\mathcal{H}_j\}\notag\\
&\qquad=G((k+1)T_s|\mathbf{\Theta}_j)-G(kT_s|\mathbf{\Theta}_j),
\label{eq:CDFG}
\end{align}
where $G(\cdot|\mathbf{\Theta}_j)$ is the CDF function for gamma distribution with parameters $\alpha_j$ and $\beta_j$. Applying~(\ref{eq:CDFG}) to~(\ref{eq:ENHJ}) we have
\begin{align}
&\mathbb{E}\left\{\left\lfloor\frac{T_{\text{on}}}{T_s}\right\rfloor\right\}
=\sum\limits_{k=1}^{\infty}k\Pr\left\{\left\lfloor\frac{T_{\text{on}}}{T_s}\right\rfloor=k|\mathcal{H}_j\right\}\nonumber\\&=\lim_{L\rightarrow\infty}\sum\limits_{k=1}^Lk[G((k+1)T_s|\mathbf{\Theta}_j)-G(kT_s|\mathbf{\Theta}_j)] \notag\\
&=\lim_{L\rightarrow\infty}(L+1)G((L+1)T_s|\mathbf{\Theta}_j)-\sum\limits_{k=1}^{L+1}G(k T_s|\mathbf{\Theta}_j)\\
&=\lim_{L\rightarrow\infty}-L\frac{\Gamma(\alpha_j,(L+1)\beta_j T_s)}{\Gamma(\alpha_j)}+\sum\limits_{k=1}^{L}\frac{\Gamma(\alpha_j,k\beta_j T_s)}{\Gamma(\alpha_j)}
\label{eq:ENtemp}\\
&=\sum\limits_{k=1}^{\infty}\frac{\Gamma(\alpha_j,k\beta_j T_s)}{\Gamma(\alpha_j)},
\label{eq:ENHJS}
\end{align}
by applying $G(kT_s|\mathbf{\Theta}_j)=\frac{\Gamma(\alpha_j)-\Gamma(\alpha_j,k\beta_j T_s)}{\Gamma(\alpha_j)}$, and the left hand part in~(\ref{eq:ENtemp}) can be shown to be zero by L'Hopital's rule. Then we introduce the following Lemma as a step to prove~(\ref{eq:ENHJS}) converges.
\begin{lemma}
\begin{align}
\int_0^{\infty}\Gamma(\alpha_j,k\beta_j T_s)dk=\frac{\alpha_j\Gamma(\alpha_j)}{\beta_j T_s}.
\label{eq:gammaint}
\end{align}
\end{lemma}
\begin{IEEEproof}
We can easily prove it by applying the change of variable technique.
\end{IEEEproof}
Since $\Gamma(\alpha_j,k\beta_j T_s)$ is a decreasing function with respect to $k$ by definition and $\int_0^{\infty}\frac{\Gamma(\alpha_j,k\beta_j T_s)}{\Gamma(\alpha_j)}dk=\frac{\alpha_j}{\beta_j T_s}$, from the integral test, we know~(\ref{eq:ENHJS}) converges. Therefore, using~(\ref{eq:ENHJS}) we can derive the average expected number of PU traffic samples by taking the average for all possible hypotheses which results in~(\ref{eq:N}).

\section{PDF Derivation for Sum of the Gamma and Triangular Distributed Random Variables}
\label{sec:likelihood_noise}

By directly convolving the PDF of gamma distributed random variable $x$, i.e., $f_X(x|\mathbf{\Theta})$ where $\mathbf{\Theta}=(\alpha,\beta)$ with the PDF of triangular distributed random variable $\phi$, i.e., $f_\Phi(\phi)$, we have the PDF for $\tilde x=x+\phi$ as
\begin{align}
&f_{\tilde X}(\tilde x)=\int_{-\infty}^{\infty}f_X(\tilde x-x|\mathbf{\Theta})f_\Phi(x)dx\notag\\
&\!\!\!\!\!\!\!\!\!=\begin{cases}
\int_{-T_s}^{0} \frac{\beta^\alpha}{\Gamma(\alpha)}(\tilde x-x)^{\alpha-1}e^{\beta(\tilde x-x)}\\
\quad\times\left(\frac{1}{T_s^2}x+\frac{1}{T_s}\right)dx\\
\quad+\int_0^{T_s}\frac{\beta^\alpha}{\Gamma(\alpha)}(\tilde x-x)^{\alpha-1}e^{\beta(\tilde x-x)}\\
\quad\times\left(\frac{-1}{T_s^2}x+\frac{1}{T_s}\right)dx,&\text{if}~\tilde x\geq T_s,\\
\int_{-T_s}^{0} \frac{\beta^\alpha}{\Gamma(\alpha)}(\tilde x-x)^{\alpha-1}e^{\beta(\tilde x-x)}\\
\quad\times\left(\frac{1}{T_s^2}x+\frac{1}{T_s}\right)dx\\
\quad+\int_0^{\tilde x}\frac{\beta^\alpha}{\Gamma(\alpha)}(\tilde x-x)^{\alpha-1}e^{\beta(\tilde x-x)}\\
\quad\times\left(\frac{-1}{T_s^2}x+\frac{1}{T_s}\right)dx,&\text{if}~0\leq \tilde x< T_s,\\
\int_{-T_s}^{\tilde x} \frac{\beta^\alpha}{\Gamma(\alpha)}(\tilde x-x)^{\alpha-1}e^{\beta(\tilde x-x)}\\
\quad\times\left(\frac{1}{T_s^2}x+\frac{1}{T_s}\right)dx,& \text{if}~-T_s\leq \tilde x<0,\\
0,&\text{otherwise}.
\end{cases}
\label{eq:PDFYs1}
\end{align}
We now introduce the following Lemma.
\begin{lemma}
\begin{align}
&\int_a^b\frac{\beta^\alpha}{\Gamma(\alpha)}(\tilde x-x)^{\alpha-1}e^{-\beta(\tilde x-x)}\left(\frac{1}{T_s^2}x+\frac{1}{T_s}\right)dx\notag\\
&=\frac{\Gamma(\alpha+1,\beta(\tilde x-a))-\Gamma(\alpha+1,\beta(\tilde x-b))}{\Gamma(\alpha)\beta T_s^2}\notag\\
&\qquad-\frac{(\tilde x+T_s)(\Gamma(\alpha,\beta(\tilde x-a))-\Gamma(\alpha,\beta(\tilde x-b)))}{\Gamma(\alpha) T_s^2},
\label{eq:lem4a}
\end{align}
\begin{align}
&\int_a^b\frac{\beta^\alpha}{\Gamma(\alpha)}(\tilde x-x)^{\alpha-1}e^{-\beta(\tilde x-x)}\left(-\frac{1}{T_s^2}x+\frac{1}{T_s}\right)dx\notag\\
&=\frac{-\Gamma(\alpha+1,\beta(\tilde x-a))+\Gamma(\alpha+1,\beta(\tilde x-b))}{\Gamma(\alpha)\beta T_s^2}\notag\\
&\qquad+\frac{(\tilde x-T_s)(\Gamma(\alpha,\beta(\tilde x-a))-\Gamma(\alpha,\beta(\tilde x-b)))}{\Gamma(\alpha)T_s^2}.
\label{eq:lem4b}
\end{align}
\label{lemma_conv}
\end{lemma}
\begin{IEEEproof}
Expression~(\ref{eq:lem4a}) and~(\ref{eq:lem4b}) can be calculated directly from the definition of incomplete gamma function and through the integration by parts technique.
\end{IEEEproof}
Finally, applying Lemma~\ref{lemma_conv} to~(\ref{eq:PDFYs1}) we obtain~(\ref{eq:mPDF}).

\section{Derivation of Variance for $\tilde y_i^{(j,k)}$}
\label{sec:new_statistics}
\vspace{-0.2cm}

We ignore the index $i$ for notation convenience and denote $\tilde x=x+\phi$. We would like to find the variance of $\tilde y^{(j,k)}=\alpha_{j,k}\log\tilde x-\beta_{j,k}\tilde x$, where $x\sim f_j(x|\mathbf{\Theta}_j)$, $\phi\sim\Lambda(-T_s,T_s)$, and $\tilde x\sim f_j(\tilde x|\mathbf{\Theta}_j,T_s)$ given in Theorem~\ref{theorem:y}. Since $\tilde x$ can be negative, $\tilde y^{(j,k)}$ may be a complex number. Therefore we define $\tilde y^{(j,k)}\triangleq\tilde y_R^{(j,k)}+j\tilde y_I^{(j,k)}$, where $\tilde y_R^{(j,k)}=\alpha_{j,k}\log\tilde x-\beta_{j,k}\tilde x$ and $\tilde y_I^{(j,k)}=0$, if $\tilde x\geq0$, and $\tilde y_R^{(j,k)}=\alpha_{j,k}\log(-\tilde x)-\beta_{j,k}\tilde x$ and $\tilde y_I^{(j,k)}=\pi\alpha_{j,k}$, otherwise. Note the PDF of $\tilde y^{(j,k)}$ can be represented as $f_{\tilde Y}^{(j,k)}\left(\tilde y^{(j,k)}\right)=f_R\left(\tilde y^{(j,k)}\right)+jf_I\left(\tilde y^{(j,k)}\right)$, where $f_R(\cdot)$ and $f_I(\cdot)$ are the PDFs with respect to the real part and imaginary part of $\tilde y^{(j,k)}$. Likewise, the variance for $\tilde y^{(j,k)}$, i.e., $\tilde\sigma_{j,k}^{2}$, is the sum of the variance of its real part $\tilde\sigma_{R,j,k}^2$ and imaginary part $\tilde\sigma_{I,j,k}^2$. 

First we calculate the variance of the imaginary part. Noting that the first and the second moment for $\tilde y_I^{(j,k)}$, which are $\mathbb{E}\left\{\tilde y_I^{(j,k)}\right\}=\pi\alpha_{j,k}\mathbb{E}\left\{\tilde x<0\right\}=\pi\alpha_{j,k}\int_{-\infty}^{0}f_j(\tilde x|\mathbf{\Theta}_j,T_s)d\tilde x$ and $\mathbb{E}\left\{\left(\tilde y_I^{(j,k)}\right)^2\right\}=\pi^2\alpha_{j,k}^2\mathbb{E}\{\tilde x<0\}=\pi^2\alpha_{j,k}^2\int_{-\infty}^{0}f_j(\tilde x|\mathbf{\Theta}_j,T_s)d\tilde x$, respectively, we can derive $\tilde\sigma_{I,j,k}^2$. The variance for the real part can be obtained through $f_R(\tilde y^{(j,k)})$. Using Lemma~\ref{lamme_inverse} and observing that there may be at most three solutions to $\tilde y_R^{(j,k)}=h(\tilde x)$, we can derive the PDF for the real part of $\tilde y^{(j,k)}$ as 
\vspace{-0.4cm}
\begin{subequations}
\begin{align}
f_R\left(\tilde y^{(j,k)}\right)&=
I(\alpha_{j,k})I(\beta_{j,k})\left\{\left|\frac{W\left(0,e^{C^{(j,k)}}\right)}{\beta_{j,k}\left(1+W\left(0,e^{C^{(j,k)}}\right)\right)}\right| f_j\left(-\frac{\alpha_{j,k}}{\beta_{j,k}}W\left(0,e^{C^{(j,k)}}\right)|\mathbf{\Theta}_j\right)\right.\nonumber\\
&\qquad \left.+I(\eta-\tilde y^{(j,k)})\left[ \left|\frac{W\left(0,e^{B^{(j,k)}}\right)}{\beta_{j,k}\left(1+W\left(0,e^{B^{(j,k)}}\right)\right)}\right| f_j\left(-\frac{\alpha_{j,k}}{\beta_{j,k}}W\left(0,e^{B^{(j,k)}}\right)|\mathbf{\Theta}_j\right)\right.\right.\nonumber\\
&\qquad \left.\left.+\left|\frac{W\left(-1,e^{B^{(j,k)}}\right)}{\beta_{j,k}\left(1+W\left(-1,e^{B^{(j,k)}}\right)\right)}\right| f_j\left(-\frac{\alpha_{j,k}}{\beta_{j,k}}W\left(-1,e^{B^{(j,k)}}\right)|\mathbf{\Theta}_j\right) \right]\right\}\\
&\qquad+I(\alpha_{j,k})I(-\beta_{j,k})\left\{\left|\frac{W\left(0,e^{B^{(j,k)}}\right)}{\beta_{j,k}\left(1+W\left(0,e^{B^{(j,k)}}\right)\right)}\right| f_j\left(-\frac{\alpha_{j,k}}{\beta_{j,k}}W\left(0,e^{B^{(j,k)}}\right)|\mathbf{\Theta}_j\right)\right.\nonumber\\
&\qquad\left.+ I(\eta-\tilde y^{(j,k)})\left[\left|\frac{W\left(0,e^{C^{(j,k)}}\right)}{\beta_{j,k}\left(1+W\left(0,e^{C^{(j,k)}}\right)\right)}\right| f_j\left(-\frac{\alpha_{j,k}}{\beta_{j,k}}W\left(0,e^{C^{(j,k)}}\right)|\mathbf{\Theta}_j\right)\right.\right.\nonumber\\
&\qquad\left.\left. + \left|\frac{W\left(1,e^{C^{(j,k)}}\right)}{\beta_{j,k}\left(1+W\left(1,e^{C^{(j,k)}}\right)\right)}\right| f_j\left(-\frac{\alpha_{j,k}}{\beta_{j,k}}W\left(1,e^{C^{(j,k)}}\right)|\mathbf{\Theta}_j\right) \right]\right\}\\
&\qquad+I(-\alpha_{j,k})I(\beta_{j,k})\left\{\left|\frac{W\left(0,e^{B^{(j,k)}}\right)}{\beta_{j,k}\left(1+W\left(0,e^{B^{(j,k)}}\right)\right)}\right| f_j\left(-\frac{\alpha_{j,k}}{\beta_{j,k}}W\left(0,e^{B^{(j,k)}}\right)|\mathbf{\Theta}_j\right)\right.\nonumber\\
&\qquad\left. + I(\tilde y^{(j,k)}-\eta)\left[ \left|\frac{W\left(0,e^{C^{(j,k)}}\right)}{\beta_{j,k}\left(1+W\left(0,e^{C^{(j,k)}}\right)\right)}\right| f_j\left(-\frac{\alpha_{j,k}}{\beta_{j,k}}W\left(0,e^{C^{(j,k)}}\right)|\mathbf{\Theta}_j\right)\right.\right.\nonumber \\
&\qquad\left.\left. + \left|\frac{W\left(1,e^{C^{(j,k)}}\right)}{\beta_{j,k}\left(1+W\left(1,e^{C^{(j,k)}}\right)\right)}\right| f_j\left(-\frac{\alpha_{j,k}}{\beta_{j,k}}W\left(1,e^{C^{(j,k)}}\right)|\mathbf{\Theta}_j\right) \right]\right\}\\
&\qquad+I(-\alpha_{j,k})I(-\beta_{j,k})\left\{\left|\frac{W\left(0,e^{C^{(j,k)}}\right)}{\beta_{j,k}\left(1+W\left(0,e^{C^{(j,k)}}\right)\right)}\right| f_j\left(-\frac{\alpha_{j,k}}{\beta_{j,k}}W\left(0,e^{C^{(j,k)}}\right)|\mathbf{\Theta}_j\right)\right.\nonumber\\
&\qquad\left.+ I(\tilde y^{(j,k)}-\eta)\left[ \left|\frac{W\left(0,e^{B^{(j,k)}}\right)}{\beta_{j,k}\left(1+W\left(0,e^{B^{(j,k)}}\right)\right)}\right| f_j\left(-\frac{\alpha_{j,k}}{\beta_{j,k}}W\left(0,e^{B^{(j,k)}}\right)|\mathbf{\Theta}_j\right)\right.\right.\nonumber\\
&\qquad\left.\left. + \left|\frac{W\left(-1,e^{B^{(j,k)}}\right)}{\beta_{j,k}\left(1+W\left(-1,e^{B^{(j,k)}}\right)\right)}\right| f_j\left(-\frac{\alpha_{j,k}}{\beta_{j,k}}W\left(-1,e^{B^{(j,k)}}\right)|\mathbf{\Theta}_j\right) \right]\right\}
\label{eq:ynfPDF}
\end{align}
\end{subequations}
where $B^{(j,k)}$ is defined as in Appendix~\ref{sec:pdf} replacing $y^{(j,k)}$ with $\tilde y^{(j,k)}$, $I(\cdot)$ is defined in Appendix~\ref{sec:pdf}, $C^{(j,k)}\triangleq\frac{\tilde y^{(j,k)}}{\alpha_{j,k}}-\log\left(\frac{\alpha_{j,k}}{\beta_{j,k}}\right)$, $\eta=\alpha_{j,k}\log\left(\left|\frac{\alpha_{j,k}}{\beta_{j,k}}\right|\right)-\alpha_{j,k}$. Therefore we can obtain $\tilde\sigma_{R,j,k}^2$ by~(\ref{eq:ynfPDF}).

\section*{Acknowledgments}

The authors would like to thank Prof. Venugopal\,V. Veeravalli and Prof.\,Alexander\,G. Tartakovsky for insightful discussions related to the MSPRT classifier.


\end{document}